\documentstyle[11pt,aaspp4,flushrt]{article}

\newcommand{\mum}{$\mu$m}
\newcommand{\cm}{cm$^{-3}$}

\newcommand{\hii}{H{\small II}}
\newcommand{\uchii}{UC~H{\small II}}
\newcommand{\heii}{He{\small II}}
\newcommand{\nee}{$n_{\rm e}$}
\newcommand{\nc}{$N'_{\rm c}$}
\newcommand{\ncuc}{$N'_{\rm c,UC}$}
\newcommand{\ncc}{$N'_{\rm c,C}$}
\newcommand{\nce}{$N'_{\rm c,E}$}

\newcommand{\degr}{$^\circ$}

\newcommand{\kms}{km~s$^{-1}$}
\newcommand{\vlsr}{$v_{\rm LSR}$}

\newcommand{\mhii}{$M_{\rm HII}$}
\newcommand{\msol}{$M_\odot$}

\newcommand{\tl}{$T_{\rm L}$}

\newcommand{\te}{$T_{\rm e}^*$}

\newcommand{\delv}{$\Delta v$}

\newcommand\hrrl{{H76$\alpha$}}
\newcommand\herrl{{He76$\alpha$}}

\begin{document}


\title{Radio Continuum and Recombination Line Study of \\ 
\uchii\ Regions with Extended Envelopes}

\author{Kee-Tae Kim}
\affil{Department of Astronomy, Seoul National University, Seoul
151-742, Korea; kimkt@astro.snu.ac.kr}
\and 
\author{Bon-Chul Koo}
\affil{Astronomy Program, SEES, Seoul National University, Seoul
151-742, Korea; koo@astrohi.snu.ac.kr}


\begin{abstract}

We have carried out 21 cm radio continuum observations of 
16 \uchii\ regions using the VLA (D-array) 
in a search for associated extended emission.
We have also observed \hrrl\ recombination line towards all the sources
and \herrl\ line at the positions with strong \hrrl\ line emission.
The \uchii\ regions have simple morphologies and 
large ($\gtrsim$10) ratios of single-dish to VLA fluxes.
We detected extended emission towards all the sources. 
The extended emission consists of one to several 
compact ($\sim$1$'$ or 0.5$-$5~pc) components and 
a diffuse extended (2$'$$-$12$'$ or 4$-$19~pc) envelope. 
All the \uchii\ regions but two are located in the compact components,
where the \uchii\ regions always correspond to their peaks. 
The compact components with \uchii\ regions
are usually smaller and denser than those without \uchii\ regions.
For individual sources, we derive the spectral types (O7$-$O4)
of the ionizing stars and the fractions of
UV photons absorbed by dust within the nebulae,
which are significantly different from previous estimates based on 
the \uchii\ regions alone.

Our recombination line observations show that
the ultracompact, compact, and extended components
have approximately the same velocity in the individual sources 
with one exception (G25.72+0.05), 
implying that they are physically associated.
The compact components in each object appear to ionized by separate
sources, while the \uchii\ regions 
and their associated compact components are likely to be ionized 
by the same sources on the basis of the morphological relations
mentioned above.
This suggests that almost all of the observed \uchii\ regions
are not `real' \uchii\ regions
and that their actual ages are much greater than their dynamical age
($\lesssim$10$^4$~yr).
We find that most of simple \uchii\ regions previously known
have large ratios of single-dish to VLA fluxes, similar to our sources.
Therefore, the `age problem' of \uchii\ regions does not seem to be 
as serious as earlier studies argued.
We present a simple model in which the coexistence of the ultracompact, 
compact, and extended components for a long ($>$10$^5$~yr) time
is easily explained by combining the Champagne flow model 
with the hierarchical structure of massive star-forming regions. 
The well-known relation between the density and diameter of \hii\ regions, 
$n_{\rm e} \propto$~D$^{-1}$, is a natural 
consequence of the hierarchical structure according to our model.  
We discuss some individual sources.

\end{abstract}

\clearpage
%
%
\section{Introduction}

Ultracompact (UC) \hii\ regions are very small ($\lesssim$0.1~pc) and 
dense ($\gtrsim$$10^4$~cm$^{-3}$) ionized regions
(Wood $\&$ Churchwell 1989a, hereafter WC89;
Kurtz, Churchwell, $\&$ Wood 1994, hereafter KCW).
They are bright radio and infrared sources
and seem to be still embedded in the parental molecular clouds. 
Thus these objects have been thought to represent 
an early evolutionary stage of massive stars. 
One of the most interesting problems related to them is that 
too many \uchii\ regions are observed in the Galaxy,
i.e., the ``age problem'':
The number of \uchii\ regions is greater by about an order of magnitude 
than expected from other indicators of massive star formation rate 
based on their dynamical age (WC89; Wood $\&$ Churchwell 1989b).
Hence, it has been considered that \hii\ regions should
remain in the ultracompact phase for a much longer ($\gtrsim$10$^5$~yr) 
period than their sound crossing time ($\lesssim$10$^4$~yr) 
in a uniform medium.
Several different models were proposed to resolve this problem: 
the infall model, photo-evaporating disk model, pressure-confined model,
and stellar-wind supporting bow shock model (see Churchwell 2000 and
references therein). These models 
present various mechanisms which constrain
\hii\ regions to remain longer in the ultracompact phase. 

On the other hand, 
there is some observational evidence to suggest that \uchii\ regions
are {\it not} as young as they appear. The ultracompactness of \uchii\
regions might arise because interferometric observations are 
insensitive to structures larger than a certain extent depending
on the baseline range.
Using the Very Large Array (VLA) D-array, 
which is sensitive to large ($\sim$15$'$) structures,
Koo et al. (1996) found that G5.48$-$0.24, known as an \uchii\ region,
is actually a large \hii\ region complex 
that contains structures of diverse length scales ranging from 
0.04 to 40~pc: It has an ultracompact core, a compact core, 
an extended halo, and a diffuse large plateau. 
The presence of the extended envelope could be direct evidence
supporting that the \uchii\ region is much older than 10$^4$~yr
(see \S~5 for more details).
It also implies that there are much more ionizing photons than 
what would have been estimated from the ultracompact core alone.
A few other groups have also noted the importance of extended emission around 
\uchii\ regions in understanding the nature of \hii\ regions (Garay et al.
1993; Kurtz et al. 1999). For example, Kurtz et al. (1999) detected
extended emission 
towards 12 out of 15 randomly selected sources
from the KCW survey, and suggested on the basis of
its morphology that the extended emission may be directly
connected with \uchii\ regions in about half (8/15) of the sources.
However, the physical relationship between \uchii\ regions
and their extended envelopes is still poorly understood.

In this study, we undertook VLA radio continuum 
observations of 16 \uchii\ regions in search of 
associated extended emission and 
detected diffuse extended emission in all the sources.
A preliminary version of these results was presented by Kim $\&$ Koo (1996).
We also made \hrrl\ and \herrl\ radio recombination line (RRL) observations 
in order to study the physical association between the \uchii\ regions 
and their extended envelopes. 
The observations are described in \S~2 and
the results are presented in \S~3.
We investigate the relationship between the parameters of 
the \uchii\ regions and their envelopes in \S~4 
and discuss the physical association between the two,
the origin of extended envelopes, and their implication in resolving the age
problem in \S~5. 
Some individual sources are discussed in \S~6
and our main conclusions are summarized in \S~7.

\section{Observations}

\subsection{21~cm Radio Continuum}

We have carried out radio continuum observations of 16 \uchii\ regions with
$\sim$40$''\times$20$''$ resolution at 21~cm (Table 1). 
The observations were made 
during 1995 February using the VLA
of the National Radio Astronomy Observatory\footnote{
The National Radio Astronomy Observatory is operated by
Associated Universities, Inc., under cooperative agreement with 
the National Science Foundation}
(NRAO) in the DnC hybrid configuration.
So our observations are sensitive to structures smaller than 15$'$.
We have selected 16 \uchii\ regions from the WC89 catalog (Table 2).
These \uchii\ regions are ones with simple morphology and 
large ($\gtrsim$10) ratios of single-dish to VLA fluxes. 
Here simple morphology means that 
they each are not resolved into two or more clearly distinct components.
This requirement is to avoid possible confusion by nearby \uchii\
regions in exploring the association between an \uchii\ region and
extended emission. 
They consist of one shell-type, two spherical, three core-halo type,
four cometary, and six irregular (or multiply peaked) \uchii\ regions.
In order to improve the {\it u-v} coverage,
each source was observed at several hour angles.
The flux density was referred to 3C~286, which was assumed to have 
flux densities of 15.03 Jy (at 1.385 GHz) and 
14.64 Jy (at 1.465 GHz). The data were edited, calibrated, imaged, and
cleaned following the standard VLA procedures
in the Astronomical Image Processing System (AIPS). 
Self-calibration was accomplished for several appropriate sources 
to reduce the rms noise level of their final images.

\subsection{\hrrl\ and \herrl\ Radio Recombination Lines}

We observed \hrrl\ RRLs,
which have a rest frequency of 14,689.99 MHz, 
toward all of the objects in our sample. The observations were undertaken 
with the 140 foot (43 m) telescope of the NRAO in 1997 February and
June. The telescope has an angular resolution (FWHM) of about 
2$'$ and a main beam efficiency of 0.58 at the observing frequency. 
In order to diminish the effects of standing waves between the telescope
and receiver, we observed with focus alternately displaced
$\pm$$\lambda$/8 from its peak value.
Both circular polarizations were observed simultaneously using two 1024 channel 
autocorrelators with 40~MHz bandwidth each, and thus the velocity 
resolution was 1.59~\kms\ after Hanning
smoothing. Each spectrum was obtained by integrating typically for 20 minutes
using frequency switching. The system temperature varied in the range 
40$-$60~K during the observing sessions depending on 
weather conditions and elevation of the
source. We mapped G5.89$-$0.39 (16$'$$\times$10$'$) and 
G5.97$-$1.17 (18$'$$\times$14$'$) in full-beam spacing.
For the other sources, we made observations at 1$-$13 positions 
including the peak positions in each envelope (see Figs. 1$a-$1$p$). 
\herrl\ (14,695.97~MHz) RRLs were also observed 
at a total of 18 positions in 6 sources (see Table 5). 
These positions are ones where strong \hrrl\ line emission was detected.

\section{Results}

\subsection{21~cm Radio Continuum}

We detected extended emission in all our sources. 
Figures 1$a-$1$p$ show their images.
The extended envelopes are from $2.'1 \times 1.'8$ to 
$14.'5 \times 10.'7$ in angular extent, and their morphologies range from a
relatively simple structure with a single peak 
(e.g., G8.14+0.23, G23.71+0.17, and G23.96+0.15)
to a fairly complicated structure with several compact
components (e.g., G10.15$-$0.34, G10.30$-$0.15, and G29.96$-$0.02).
All the \uchii\ regions except for G23.46$-$0.20 and G25.72+0.05 are located  
at the peak positions. Both of the latter are spherical \uchii\ regions  
lying near the edge of extended \hii\ regions. 
If we adopt the distances given by WC89 (Table 2),
the geometric mean diameters range from 0.56~pc to 5.68~pc for the
compact components and from 3.8~pc to 18.7~pc for the extended envelopes.
Based on their linear sizes (and physical parameters),
the compact components would have been classified as compact or
dense \hii\ regions while the extended components would have been
categorized as classical \hii\ regions (Habing \&\ Israel 1979).
Table 3 presents the basic observed parameters of our sources.
In the first column of this table,
A$-$C represent the brightest three compact components in order of decreasing
peak brightness.
The first entry for each source refers to the extended component
including \uchii\ region and compact component(s).

In the sources with two or more compact components, the \uchii\ regions
tend to lie in the smallest and densest ones.
This is shown in Figure 2, where we plot angular and linear sizes versus
distance for the compact components. 
Most (11/14) of the compact components with \uchii\ regions are
smaller than 2~pc, whereas all but two without \uchii\ regions are larger.
The average diameter (1.6~pc) of the compact components 
with \uchii\ regions is about half of that of those 
without \uchii\ regions (3.5~pc). 
On the other hand, there seems to be no relationship between the
morphologies of \uchii\ region and its extended envelope.

Assuming that the \hii\ region is spherically symmetric, optically thin, 
homogeneous, dust-free, and ionization-bounded,
we derived the electron number density \nee, emission measure $EM$, 
excitation parameter $U$, and mass \mhii\ of the ionized gas 
from the observed integrated flux densities 
(Schraml $\&$ Mezger 1969; Panagia $\&$ Walmsley 1978).
We also estimated the Lyman continuum photon flux
\nc\ required to ionize each source
using the formulae of Rubin (1968).
Here we used 
the electron temperatures given by 
Downes et al. (1980) and Wink et al. (1982). 
Table 4 summarizes the results. 
The table also lists the spectral type of the ionizing star
of each source (Panagia 1973). This assumes that a
single zero-age main sequence (ZAMS) star is responsible for 
the ionization.
The spectral types of the compact components with \uchii\ regions are 
approximately three subclasses earlier than those
measured from \uchii\ regions alone in WC89, which are given in parentheses.
If a source is produced by a stellar cluster,
rather than by a single star,
then the most luminous star in the cluster would be about two subclasses
later than the equivalent single-star spectral type (see, e.g., KCW).
We determined the fraction of ultraviolet (UV) photons absorbed 
by dust within the ionized gas, $f_d$=1$-$\nc/$N^*_c$, 
where $N^*_c$ is the total rate of Lyman
continuum photons of the ionizing star (Table 4).
In this calculation, we adopted the value of $N^*_c$ from WC89,
which was derived primarily from the $IRAS$ data under the assumption 
that a single star provides the entire infrared luminosity.
We used the value of the associated (or the nearest) compact component for \nc, 
considering the low ($\sim$4$'$) resolution of $IRAS$ observations.
Our values of $f_d$ might be overestimates
because the sizes of the compact components are frequently smaller than
the $IRAS$ beamsize.
In the majority (11/16) of our sources, nevertheless,  
the estimated fractions ($<$60\%) are significantly less than 
those ($\gtrsim$80\%) determined by WC89 and KCW
for most \uchii\ regions in their catalogs. 
This discrepancy may be mainly due to the contribution from 
extended emission around \uchii\ regions.
Accordingly,
the dust absorption of stellar UV radiation within the \hii\ regions 
seems to be less important than suggested by previous studies.

\subsection{\hrrl\ and \herrl\ Radio Recombination Lines}

We detected \hrrl\ line emission in all the sources except G12.43$-$0.05,
which was the weakest radio continuum source in our sample.
Figure 3 shows a sample from our RRL spectra at 
G10.15$-$0.34 (+05.$^{\!\!\rm s}$2, +00$'$ 10$''$),
where the strongest \hrrl\ line emission was detected.
We performed a least-squares fit of each spectrum to a single Gaussian profile.
Table 5 gives the Gaussian parameters of the \hrrl\ lines. 
In this table, columns
(2)$-$(5) are the equatorial coordinates, the central velocity, 
the line intensity, and the line width (FWHM), respectively. 
The equatorial coordinates are given as offsets with respect to the
position of \uchii\ region.
Errors quoted are statistical $\pm$1~$\sigma$ errors in the fit
only.
The central velocities of our RRLs detected toward \uchii\ regions are in
good agreement with those of previous single-dish observations
(Downes et al. 1980; Wink et al. 1982; Lockman 1989).
We can find no significant velocity difference
between the spectrum observed toward each \uchii\ region and
the average spectrum of its extended envelope 
with one exception (G25.72+0.05).
This indicates that the two components are physically associated.
In the case of G25.72+0.05, \hrrl\ line emission was observed at 98~\kms\
toward the \uchii\ region while it was detected at 54~\kms\ 
in the nearby compact component,
suggesting that they constitute a chance coincidence in the line of sight.
In seven sources (G5.89$-$0.39, G5.97$-$1.17, G10.15$-$0.34,
G10.30$-$0.15, G12.21$-$0.10, G23.46$-$0.20, and G29.96$-$0.02),
on the other hand, the ionized gas in some part of the envelope is blue- or
redshifted by 3$-$10~\kms\ with respect to the compact component
associated with an \uchii\ region. 
For instance, the eastern part of G10.15$-$0.34
is blueshifted from its central region by
$\sim$10~\kms\ (Fig. 4). We will discuss in detail the velocity
structure of individual sources in \S~6. 
The \hrrl\ line width  is usually larger 
in the compact components, especially in the ones with \uchii\ regions,
than in the extended envelopes for the resolved sources 
(see \S~6.1 and \S~6.2).
This trend probably occurs because turbulent motions,
including large-scale expansion or contraction motions, 
are larger in the compact components than in the extended envelopes, 
as proposed by Garay \&\ Rodr{\'\i}guez (1983).

\herrl\ line emission was detected at 10 out of the 18 observed 
positions (Table 6). The singly ionized helium abundance (by number) Y$^+$ 
was estimated at these positions from 
the ratio of the \herrl\ to \hrrl\ line integrals.
The derived values (0.063$-$0.080) for G10.30$-$0.15, G23.46$-$0.20,
and G29.96$-$0.02 are comparable to the average value (0.074$\pm$0.003) 
in our Galaxy (Shaver et al. 1983), 
while the values (0.016$-$0.049) for G5.89$-$0.39, G5.97$-$1.17,
and G10.15$-$0.34 are significantly lower.
The low value of Y$^+$ implies that 
either the sizes of \heii\ regions are significantly smaller than those of
\hii\ regions or the He abundance Y is low at the observed positions.
For G5.89$-$0.39 and G5.97$-$1.17,
the observed low value is likely to be due to the low effective
temperature of the ionizing stars (O7) (Mezger 1980).
Additionally these two objects are at galactocentric distances of
6$-$7~kpc where the metal abundance is known to be significantly higher
than in the solar neighborhood (Shaver et al. 1983). The high
metallicity could make the sizes of \heii\ regions smaller.
This is not true for G10.15$-$0.34, which has ionizing 
stars of spectral type O5 or earlier.
Such stars emit hard enough photons to singly ionize
helium in the entire HII region. 
This view is supported by the fact that
normal values were observed for Y$^+$ in G10.30$-$0.15,
G23.46$-$0.20, and G29.96$-$0.02, which have ionizing stars of O5 or O5.5. 
We can also exclude the possibility that
Y is low in the region because the value ($\geq$0.063)
is approximately normal in G10.30$-$0.15, which is located in the same
molecular cloud (see \S~6.3).
A possible explanation for the low Y$^+$ observed in G10.15$-$0.34 is that 
helium ionizing photons are preferentially absorbed by dust in the
nebula as in Sgr~B2 (e.g., Chaisson, Lichten, \&\ Rodr{\'\i}guez 1978).


\section{Relationship between the Parameters of \uchii\ Regions \\
and Their Extended Envelopes }

\subsection{Velocity}

What is the physical relationship between the ultracompact, compact,
and extended components in each source?
According to our RRL observations,
the compact component(s) and extended envelope appear to be physically 
associated in the individual sources.
On the other hand,
our measurements of the \hrrl\ line emission toward \uchii\ regions were 
likely to be dominated by their associated compact components,
rather than that due to the \uchii\ regions themselves,
since our beam size ($\sim$2$'$) is much larger than the typical 
size ($\lesssim$10$''$) of the \uchii\ regions. 
In order to investigate the physical relationship between \uchii\ regions
and their associated compact components, therefore, 
we compare our \hrrl\ line data with the RRL and molecular line data of 
\uchii\ regions taken from the literature (Table 7).
In this comparison, we do not include G12.43$-$0.05,
in which the \hrrl\ line was not detected. 
G23.46$-$0.20 and G25.72+0.05 are also excluded 
because they have no associated compact components.
Figure 5 exhibits the results. 
On the whole, the central velocities of the two components 
are in rough agreement and there is no systematic trend between 
the two parameters. 
For 5 \uchii\ regions,
our results are compared with those of high-resolution RRL observations
(Andrews et al. 1985; Churchwell et al. 1990b; Afflerbach et al. 1996).
The central velocity of the \uchii\ region is equal to that of 
its associated compact component within 2~\kms\ for 
G6.55$-$0.10, G12.21$-$0.10, and G29.96$-$0.02, 
and within 6~\kms\ for 
G5.89$-$0.39 and G23.96+0.15. 
For the other \uchii\ regions,
our \hrrl\ lines are compared with their associated NH$_3$ or CS (7$-$6) 
lines (Churchwell et al. 1990a; Cesaroni et al. 1992; Plume et al. 1992).
These molecular lines trace the densest part of a cloud, and their
velocities should be close to those of \uchii\ regions
(e.g., Forster et al. 1990).
We could not find any significant ($\gtrsim$6~\kms) difference 
between the two velocities except for G23.71+0.17, for which 
the difference is $\sim$10~\kms. This large velocity difference is probably 
due to a relative motion between the ionized and molecular gas. 
Table 7 also presents the average  difference between 
the velocities of the ultracompact and compact components,
$<v_{\rm LSR,UC}-v_{\rm LSR,C}>$, for each line emission.

\subsection{Ionizing Photons}

Figure 6 displays \ncuc/\ncc\ versus \ncc/\nce\ for 14 sources 
in which the \uchii\ region has an associated compact component.
Here \ncuc, \ncc, and \nce\ are the Lyman continuum photon fluxes 
of the ultracompact, compact, and extended components, respectively.
For \ncuc, we used the values estimated by WC89 from flux densities 
at 2~cm or 6~cm.
The ratios of \ncc/\nce\ are greater than 65\%
for all the sources with a single peak except for G5.97$-$1.17, 
which has a ratio of about 30\%. 
On the other hand,
\ncuc/\ncc\ is less than 35\% for 12 of the 14 sources.
Our values of \ncuc/\ncc\ and \ncc/\nce\ may be, respectively,
overestimated and underestimated and so should be taken as upper limits
and lower limits,
since \uchii\ regions are optically thin at 2~cm or 6~cm but optically
thick at 21~cm.
The extreme cases are G5.89$-$0.39 and G29.96$-$0.02, for which
\ncc\ is equal to or less than \ncuc.


\section{Discussion}

\subsection{Physical Association of \uchii\ Regions and Their Extended
Envelopes}

The \uchii\ region, compact component(s), and extended envelope in each
source have approximately the same radial velocity 
and thus the three components are likely to be physically associated 
in almost all of our sources.
Are they ionized by the same source or by separate sources?
Although they are physically associated, they can still be
ionized by separate sources, e.g., by stars of different ages in associations.
This is closely related to how massive stars form from molecular clouds
and how they destroy the natal clouds.
First, in order to examine whether the compact components in each source
are excited by the same ionizing source or not,
we estimated how much Lyman continuum photon flux the ionizing star
in the strongest compact component can provide for nearby components 
in sources with two or more compact components.
Here 
it was assumed that the strongest component is density-bounded and
the star maintains the ionization of the entire \hii\ region
(including all compact components and extended envelope).
The flux supplied depends on the solid angle subtended
by the compact component from the ionizing star.
The derived fluxes are much less (0.3\%$-$30\%)
than the required values in all the sources.
In the case of G10.15$-$0.34, for example, the ionizing source in the
eastern compact component can supply the western one with
at most 15\% of the required flux.
{\it Hence, the compact components in each source
appear to be ionized by separate sources.}

What about the compact components and their envelopes?
As stated in \S~4.2, the ratios of \ncc/\nce\ are greater than 65\%
for all the sources with a single peak except for G5.97$-$1.17. 
The surface brightness decreases monotonically from the compact
component peak to the envelope edge in these sources.
{\it Therefore,
it seems probable that the extended envelope
is entirely or at least significantly ionized by the exciting star of
the compact component.}
In the case of sources with two or more compact components,
the envelope of each compact component could have been produced
by UV photons escaping from the compact component 
in a similar manner to the sources with a single peak.
The individual envelopes might have merged to form a larger 
envelope at later times.
However, we can not exclude absolutely the possibility that 
the large envelope may have been created by separate ionizing sources, 
such as massive stars in an earlier generation 
as in the sequential star formation model (Elmegreen \& Lada 1977).

We can also ask whether the \uchii\ region and its associated compact 
component are excited by the same source or by separate sources.
As noted in \S~3.1,
the \uchii\ regions correspond to the peaks of their associated compact
components in all the sources
and the compact components with \uchii\ regions 
are usually smaller and denser
than those without \uchii\ regions. 
For three sources, i.e., G5.89$-$0.39, G6.55$-$0.10, and G29.96$-$0.02,
we were able to confirm that the \uchii\ region is still located at the peak 
of radio continuum emission in available data obtained at higher resolution
(Zijlstra et al. 1990; Andrews, Basart, \& Lamb 1985; Fey et al. 1995).
If they are ionized by independent sources, one would not expect 
these correlations.
{\it Hence 
the UC~H{\small $II$} region and its associated compact component are likely to
be excited by the same ionizing source in our samples.} 
If this is true,
the fractions of UV photons that escape from the \uchii\
regions, $f_{esc}$=1$-$\ncuc/\ncc, are estimated
to be 65\%$-$97\% for our sources. 
These are probably underestimates (see \S~4.2)

\subsection{Is the Age Problem Solved?}

The age problem of \uchii\ regions originates from the results of 
Wood \& Churchwell (1989b), who found that \uchii\ regions have peculiar
$IRAS$ 60-12~\mum\ and 25-12~\mum\ colors,
log$(F_{60} / F_{12}) \geq 1.30$ and log$(F_{25} / F_{12}) \geq 0.57$,
and evaluated the total number of \uchii\ regions in the Galaxy
to be 1650$-$3300 based on these color criteria. 
Since the $IRAS$ sample is sensitivity-limited, most of them are likely to be
embedded O stars. 
This estimate is about 10 times greater than the number derived from
the known formation rate of O stars, 
e.g., 4$\times$10$^{-2}$ O stars~yr$^{-1}$ (see Garay \& Lizano 1999), 
and the dynamical age ($\lesssim$10$^4$ yr) of \uchii\ regions.
However,
almost all of the \uchii\ regions in our sample 
have extended ($\gtrsim$ a few 1 pc) envelopes ionized by the same sources, 
which implies that their actual ages are $\gtrsim$10$^5$~yr (see next section).
If most \uchii\ regions are associated with extended emission as in our
sources, therefore, the age problem is expected to be resolved 
although we need to explain how structures with very different
length scales can coexist in an \hii\ region for a long time.

In order to address this issue, we calculated the ratios of 
single-dish ($S_{\rm SD}$) to VLA fluxes ($S_{\rm VLA}$)
for 30 \uchii\ regions with simple morphology in the WC89 catalog. 
Here we used the 11~cm flux densities observed with the Effelsberg 
100~m telescope (FWHM$\simeq$4.$\!'$3) (F$\ddot{\rm u}$rst et al. 1990). 
Most of the \uchii\ regions have large ratios (Fig. 7). For instance, 
about 70\% have ratios $>$5 or $S_{\rm VLA}$/$S_{\rm SD}$$<$0.2.
The derived ratios are likely to be underestimated,
since the \uchii\ regions are quite optically thick at 11~cm.
However, this may have little effect on our result.
One may suspect that the result is attributed to selection effects of 
the WC89 survey, because their original target list was created by
selecting strong compact sources from previous single-dish surveys. 
Thus we also determined the flux ratios for 22 simple \uchii\ regions 
in the survey of KCW,
which were chosen by the $IRAS$ color criteria of Wood \& Churchwell
(1989b), and obtained a very similar result (Fig. 7).
For comparison, 
we computed the ratios for the 15 sources of Kurtz et al. (1999),
which consist of 8 simple and 7 complex \uchii\ regions. 
About 80\% of the 12 sources with extended emission have ratios $\gtrsim$10,
while two of the 3 sources without extended emission have ratios $<$5.
{\it Consequently,
most sources known as UC~H{\small $II$} regions are likely to
be associated with extended emission, analogous to our sources.}

The $IRAS$ color criteria of Wood \& Churchwell (1989b)
were established using the \uchii\ regions in the WC89 catalog. 
So the color criteria
may select compact or extended \hii\ regions as well as \uchii\ regions.
This idea is consistent with the results of Codella, Felli, \& Natale (1994)
who found that more than half of 445 diffuse \hii\ regions are related
to $IRAS$ point sources that satisfy the color criteria.
We have investigated whether or not there is a systematic trend
between the $IRAS$ colors and $S_{\rm SD}$/$S_{\rm VLA}$.
The $IRAS$ colors may be expected to become bluer with 
increasing $S_{\rm SD}$/$S_{\rm VLA}$ 
because the color criteria for the selection of $IRAS$ point sources 
associated with diffuse \hii\ regions are known to be bluer than 
those for \uchii\ regions (Hughes \& MacLeod 1989). 
However, we could not find any apparent variation in the $IRAS$ colors 
with $S_{\rm SD}$/$S_{\rm VLA}$,
which implies that the $IRAS$ colors are not very sensitive to 
the presence of extended emission.

\subsection{Origin of Extended Envelopes}

Our observations show that
an \uchii\ region is usually surrounded by more extended 
emission, which appears to be excited by the same ionizing source.
Various models were suggested to explain the longevity of \uchii\ regions, 
as mentioned in \S~1, but none of them predicted the presence of 
extended emission around them.
We consider the origin of the extended envelopes 
of \uchii\ regions in this section.
Molecular clouds are well known to be hierarchically clumpy
at all observable length scales.
{\it Thus it is not surprising that H{\small $II$} regions, which form in 
molecular clouds, have substructures as in our sources.}
Recent high-resolution studies of massive star-forming regions 
have revealed ``molecular clumps'' and ``hot cores'' therein; 
the molecular clumps have diameters $\sim$1~pc, masses
$>$10$^3$~\msol, and densities $\sim$10$^{5}$~\cm, 
while the hot cores have diameters $\lesssim$0.1~pc, masses of
(1$-$3)$\times$10$^2$~\msol, densities $>$10$^6$~\cm, and
temperatures $\gtrsim$100~K 
(see Kurtz et al 2000 and references therein).
The signposts of massive star formation, such as \uchii\ regions and maser
sources, are commonly observed in the molecular clumps. 
The hot cores are closely associated with \uchii\ regions.
They are believed to be the sites of massive star formation
where \uchii\ regions have not yet developed (e.g., Cesaroni et al. 1994)
or remnants of the gas from which the massive stars formed
(e.g., Garay \& Rodr{\'\i}guez 1990).
The sizes of the hot cores and molecular clumps agree roughly with 
those of \uchii\ regions and their associated compact components in our
sample, respectively.  

We can first think of the possibility that
the extended envelope could be formed very soon after the central star
turns on, e.g., before the \hii\ region starts to expand dynamically.
It may depend on the density of interclump medium and 
volume filling factor of dense clumps.
According to earlier studies on the internal
structures of the molecular clumps in massive star-forming regions
(Snell et al. 1984; Stutzki et al. 1988; Stutzki \& G\"{u}esten 1990;
Plume et al. 1997), 
the interclump medium density is generally $\gtrsim$10$^4$~\cm\ and 
the volume filling factor varies between $<$0.1 and 0.5
and tends to increase with approaching the peak.
Assuming that the spectral type of ionizing star is O6,
the initial Str\"{o}gren radii of \hii\ regions 
would be $<$0.15~pc, which are much less than the typical size of 
molecular clumps. 
To our knowledge, there is no previous
detailed study on the substructures of the hot cores. 
However, both the interclump medium density and 
volume filling factor are expected to be greater in the hot cores,
since they have much higher mean densities.
Therefore, the extended envelope is likely to formed   
by the dynamical expansion of ionized gas sphere. 
The existence of \uchii\ regions without extended envelopes,
such as G23.46$-$0.20, G25.72+0.05 (\S~6.4), 
and G78.4+2.6 (Kurtz et al. 1999), also argues for this view. 

We propose a simple model in which
{\it the existence of the extended emission around UC~H{\small $II$} regions 
can be explained by combining the Champagne flow model with 
the hierarchical structure of massive star-forming regions}
(Tenorio-Tagle 1982 and references therein).
Figure 8 is a schematic representation of our model.
Here we consider the case where a massive star forms 
off-center within a hot core with $n_{\rm H_2}$$\sim$10$^7$~\cm\
and $M$$\sim$100~\msol,
which is placed in a molecular clump of $n_{\rm H_2}$$\sim$10$^5$~\cm.
The initial and final equilibrium Str\"{o}mgren radii in this case are,
respectively, about 0.0015~pc and 0.05~pc assuming  
\nc=10$^{49}$~s$^{-1}$ for the star (O6).  
A champagne flow would develop when an ionization front (of D-type) 
breaks out of the core.
The timescale that the ionization front needs to reach the core's edge
depends on the depth of the exciting star inside the core.
Assuming that the depth is 0.03~pc,
the timescale will be $\sim$10$^4$~yr from the expansion
law for \hii\ regions in a uniform medium (Spitzer 1968). 
If \nc\ is significantly less than 10$^{49}$~s$^{-1}$ and/or
$n_{\rm H_2}$ is greater than 10$^7$~\cm, the \hii\ region can reach
pressure equilibrium without breaking through the core's boundary
as in the pressure-confined model of \uchii\ regions
(De Pree, Rodr{\'\i}guez, $\&$ Goss 1995; Garc{\'\i}a-Segura $\&$ Franco 1996).
The supersonic gas flow drives a strong isothermal shock moving away 
from the core and a rarefaction wave traveling toward the exciting star. 
The velocity of the champagne flow, which is determined by  
the pressure contrast at the edge, could increase up to 30~\kms. 
According to Whitworth (1979) who
estimated the efficiency with which O stars disperse their parental
molecular clouds, the erosion timescale can be calculated from
the following formula

\begin{equation}
t~\approx~3 \times 10^5 ~ 
{\Bigl( \frac{M}{10^2~M_\odot} \Bigr)}^{7/9}~
{\Bigl(\frac{N'_{\rm c}}{10^{49}~{\rm s}^{-1}} \Bigr)}^{-4/9}~
{\Bigl(\frac{n_{\rm H_2}}{10^7~{\rm cm}^{-3}} \Bigr)}^{1/9}~~~~({\rm yr}).
\end{equation}

\noindent
If this result is adopted,
the lifetime of the hot core would be about 3$\times$10$^5$~yr. 
Meanwhile, the \hii\ region continues to be ultracompact within the hot
core 
while it could grow up to a few 1~pc outside the core.
Another champagne flow would develop
when the ionization front crosses the edge of the molecular clump.
Lyman continuum photons that escape from the clump ionize the surrounding 
region with a lower density, which forms a more extended \hii\ region.

We could find the champagne flows in 
several sources \S~3.2 based on the velocity gradient of \hrrl\ line emission
and/or the morphology of radio continuum image (see also \S~6).
This is compatible with expectation of our model.
It is difficult to detect the champagne flows in all the sources
because of the large line width and non-Gaussian profile of 
\hrrl\ lines as well as various viewing angles.
We may not expect to obtain direct evidence for the large velocity 
gradients on length scales of molecular clumps or smaller owing to 
the low resolution and coarse sampling of our RRL observations.
However, the larger line width of RRLs in the compact components
noted earlier is consistent with our model.
In the light of the model presented here, among our sources
G5.97$-$1.17 and G8.14+0.23 are good examples of
\hii\ regions observed at viewing angles of about 0\degr\ and 90\degr,
respectively (see \S~6.2 and \S~6.6.2).

\subsection{Density versus Size Relationship}
 
For UC and compact \hii\ regions,
\nee\ was found to be on the average proportional to D$^{-1}$
(Garay et al. 1993; Garay \& Lizano 1999).
Figure 9 compares electron density with diameter for the compact 
and extended components of our sources.
There is a fairly good correlation between the two parameters.
A least-squares fit yields \nee~=~630~D$^{-0.93\pm0.05}$.
Figure 9 also displays the relation between \nee\ and D
for more compact \hii\ regions in the catalogs of WC89, KCW,
and Garay et al. (1993).
We performed a least-squares fit to all the data points in the plot 
and obtained
\nee~=~790~D$^{-0.99\pm0.03}$.

If an \hii\ region evolves with a constant flux of ionizing photons,
\nee\ would be roughly proportional to D$^{-1.5}$.
In view of this, the previous studies suggested two possible explanations 
for the shallower power law:
either \uchii\ regions are excited by stars with lower luminosities than
those exciting compact \hii\ regions 
or a part of \uchii\ regions in the surveys of WC89 and KCW
are externally ionized dense clumps within inhomogeneous \hii\ regions.
According to our model in \S~5.3, however, the observed D$-$\nee\ relationship
is not a manifestation of the evolutionary effect 
but mainly of the variation in the ambient density;
more compact \hii\ regions are located in denser parts of molecular clouds. 
In this regard, it is natural to find a D$-$\nee\ relationship which
is similar to the D$-$$n_{\rm H_2}$ relationship of molecular clouds,
e.g., $n_{\rm H_2} \propto$~D$^{-1.1}$ (e.g., Larson 1981).
There could be some evolutionary effects for larger \hii\ regions, 
but Figure 9 shows that the effect does not make a significant contribution 
to the relationship for \hii\ regions with D$\lesssim$20~pc.


\section{Discussion of Some Individual Sources}

\subsection{G5.89$-$0.39}

G5.89$-$0.39, also known as W28~A2(1), is situated in W28~A2,
which is $\sim$50$'$ south of the W28 supernova remnant (SNR).
This \uchii\ region was found to drive one of the most powerful 
($\dot P$$\simeq$0.33 \msol~\kms~yr$^{-1}$)
molecular bipolar outflows (e.g., Accord, Walmsley, \& Churchwell 1997)
and thus has been extensively studied. 
In the VLA radio continuum maps with high ($<$1$''$) resolution,
G5.89$-$0.39 appears as a $\sim$5$''$ shell (WC89) 
or a disk-like structure (Zijlstra et al. 1990).
Previous estimates of the spectral type of the ionizing star 
range from O5 to O7.
 
Our radio continuum observations show that there are two 
compact components separated by 2$'$.5 in this field:
a large (1.1~pc) central one 
and a small (0.7~pc) western one (Fig. 1$a$). 
G5.89$-$0.39 is located in the western compact component.
The two compact components are embedded in low-level emission
extending over $14' \times 9'$ (or $10.9 \times 6.8$~pc$^2$ at 2.6~kpc). 
The brightness distribution within the extended structure 
is fairly complicated.
A bright ridge, originating from the central compact component, extends
in the southwest direction. 
The surface brightness 
declines outward from the central region and then
increases again at the edge of the extended structure. 
Figures 10$a$ and 10$b$ show the distributions of the \hrrl\ line integral and
equivalent line width ($\int T_{\rm L} dv$/\tl) in this field, respectively.
The line integral peak
does not correspond to the central continuum peak but to the western 
peak with the \uchii\ region. The line width is largest
(47~\kms) at the same position, as pointed out in \S~3.2. 
This large line width is certainly attributed to the energetic outflow.  
The central velocity is slightly lower in the
compact components than in the extended envelope.

\subsection{G5.97$-$1.17}

G5.97$-$1.17 lies in the well-observed HII region M8, the Lagoon nebula. 
M8 is placed in the Sagittarius arm of the Galaxy
and contains an extremely young (about 1.5~Myr) open cluster, NGC 6530,
in the eastern part (Sung, Chun, \& Bessell 2000).
The dense region ($\sim$10$'$) of M8 in optical emission corresponds
roughly to our radio \hii\ region.
The central part (30$''$$\times$15$''$ in NS$\times$EW) of the dense region 
is known as the Hourglass nebula. 
G5.97$-$1.17 and the Hourglass, respectively, lie $\sim$3$''$ 
to the southeast and $\sim$15$''$ east of the O7~V star Herschel 36,
which is known to be a newly formed star recently emerged from 
its dust cocoon (WC89; Woolf 1961). 
It was suggested that the dense region is 
ionized mainly by Herschel 36 while the more extended nebulosity 
is excited by the other O stars in NGC~6530, 
such as 9 Sgr (O4(f)) and HD~165052 (Oe) 
(Lada et al. 1976; Elliot et al. 1984).
Based on their optical and radio observations showing that
the ionized gas is blueshifted by about 7~\kms\ with respect to
the underlying molecular cloud,
Lada et al. (1976) proposed that
the dense ionized region is on the near edge of the cloud,
similar to the Orion Nebula.  
Stecklum et al. (1998) recently suggested that G5.97$-$1.17 is not a
real \uchii\ region 
but a circumstellar disk surrounding a young star (later than B5),
which is being photoevaporated by Herschel 36.  

Our radio continuum observations reveal a single compact component
surrounded by extended emission of 14.$'5 \times 10.'7$ 
(or $8.0 \times 5.9$~pc$^2$ at 1.9~kpc) (Fig. 1$b$). 
G5.97$-$1.17 and the Hourglass nebula are located in the compact component.
Since the surface brightness declines monotonically from the \uchii\ region
to the edge of the extended envelope, the ionizing source of our radio \hii\
region is expected to be located in the compact component. 
Herschel~36 may be the best candidate.
An interesting feature is the arclike structure emanating 
from the southeastern part of the envelope. 
Figures 11$a$ and 11$b$ show maps of the \hrrl\ line integral and
equivalent line width for this field, respectively.
The distribution of the line integral is similar to that of continuum emission, 
but the line width is extraordinarily large at the bottom of 
the arclike structure.
This large velocity dispersion is likely to be a consequence of
the champagne flow,
although the velocity difference between the arc and compact component
is not large ($\sim$3~\kms).
The central velocity appears to be larger by a small amount
in the northeastern part than in the remaining regions.

\subsection{G10.15$-$0.34 and G10.30$-$0.15}

These two \uchii\ regions are located in the W31 HII region/molecular cloud 
complex. In lower-resolution radio continuum maps,
W31 comprises two strong components 
which contain G10.15$-$0.34 and G10.30$-$0.15, respectively. 
The SNR G10.0$-$0.3, 
which was identified as the
soft $\gamma$-ray repeater SGR~1806$-$20 (Kulkarni \&\ Frail 1993), 
is $\sim$9$'$ southwest of G10.15$-$0.34.

Our radio continuum observations show in the field of G10.15$-$0.34
that there are several compact components including two strong ones 
and that they are embedded in extended emission of
$10.'9 \times 6.'7$ (or $19.0 \times 11.1$~pc$^2$ at 6.0~kpc) (Fig. 1$e$).
The \uchii\ region lies in the western compact component. 
The central region is extended
in the east-west direction. Based on their 
VLA continuum observations at 5~GHz,
Woodward et al. (1984) suggested that this elongated region consists of
about 20 dense ionized clumps and that the individual clumps are HII regions
excited by separate embedded stars of spectral types B0.5$-$O6.5. 
The surface brightness declines rapidly to the west, while it decreases
slowly to the east.
As mentioned in \S~3.2,
our RRL observations strongly suggest
that the eastern protuberance may be a result of the champagne flow (Fig. 4).

In the field G10.30$-$0.15 (Fig 1$f$), we can see two compact components 
surrounded by diffuse emission extending over $12.'8 \times 4.'6$ 
(or $22.3 \times 8.0$~pc$^2$ at 6.0~kpc).
The two compact components form an ionized ridge
elongated in the NE-SW direction and the western  one contains G10.30$-$0.15. 
The diffuse envelope extends straight in the NW-SE direction
perpendicular to the central ridge. 
Such a bipolar HII region can be
produced by the champagne flow in the case where the ionizing source is located
in a thin, flat molecular cloud (Bodenheimer, Tenorio-Tagle, \& Yorke 1979).
According to our molecular line observations (Kim \&\ Koo 2000), the
distribution of the ambient molecular gas strongly supports 
the Champagne flow model. 
However, we could not observe a significant velocity
difference between the central region and the envelope. 
This is presumably because
the inclination angle of the extended structure is close to zero.
We will present the result of a detailed infrared and radio study 
on the W31 region in a forthcoming paper (Kim \&\ Koo 2000).

\subsection{G23.46$-$0.20 and G25.72+0.05}
 
These two spherical \uchii\ regions
have no associated compact components.
The extended \hii\ region associated with G23.46$-$0.20
consists of three compact components and a diffuse envelope elongated in the
east-west direction (Fig. 1$i$).
G23.46$-$0.20 is located near the edge of the extended \hii\ region.
The velocities of the northern and western parts
are lower than that of the central region by $\sim$4~\kms\ (see Table 5).
G25.72+0.05 is also located on the sky near the edge of an extended 
\hii\ region with a single compact component (Fig 1$l$)
but the velocity of G25.72+0.05 is very different from that of
the extended \hii\ region as noted in \S~3.2.
This strongly suggests that
they are a chance coincidence in the line of sight.
NH$_3$~(2,2) lines with almost same velocity (\vlsr$\simeq$100~\kms) 
as the \hrrl\ lines were detected towards G23.46$-$0.20 and G25.72+0.05 
(Churchwell et al. 1990a).
Thus these two \uchii\ regions are likely to be real \uchii\ regions.
Their exciting stars were determined to be B0 stars by WC89.
However, we can not rule out absolutely
the possibility that they may be externally ionized dense clumps, 
especially in the case of G23.46$-$0.20.

\subsection{G29.96$-$0.02}

G29.96$-$0.02 is a prototypical cometary \uchii\ region and has been
studied in detail at infrared and radio wavelengths. 
This \uchii\ region appears
as a $\sim$5$''$ diameter edge-brightened arclike structure in the VLA
maps of WC89. 
In order to explain the cometary morphology and
velocity structure of this source,
two competing models were proposed:
the bow shock model (Wood \&\ Churchwell 1991; Mac Low et al. 1991;
Afflerbach et al. 1994) and the Champagne flow model (Fey et al. 1995;
Pratap, Megeath, \&\ Bergin 1999; Lumsden \&\ Hoare 1999).
Recent near-infrared images  show that G29.96$-$0.02 lies in an
embedded cluster
(Fey et al. 1995; Pratap et al. 1999).
Watson \& Hanson (1997) restricted the exciting star to spectral classes
O5$-$O8 on the basis of its $K$-band spectrum.

Our radio continuum observations reveal 
several compact components including two strong ones
in this field (Fig. 1$n$). G29.96$-$0.02 exists in the northern compact
component from which two protuberances stretch to the east and 
to the southwest, respectively. 
Most compact components are embedded in extended emission
of $6.'3 \times 5.'2$ (or $16.5 \times 13.6$~pc$^2$ at 9 ~kpc). 
There does not exist a significant velocity difference between the compact
components and extended envelope, 
suggesting a physical association between them.

\subsection{Other Sources}

\subsubsection{G6.55$-$0.10}
G6.55$-$0.10, also known as W28~A1, is located on the sky near 
the center of the W28 SNR, which appears to be
an incomplete shell with a diameter of 
$\sim$30$'$ in our radio continuum map (Fig. 1$c$).
The distance to W28 is known to be $\sim$2~kpc (Kaspi et al. 1993),
whereas G6.55$-$0.10 seems to be located at a much greater distance of 
16.7~kpc on the basis of H$_2$CO absorption line observations 
(Downes et al. 1980).
Hence, the two objects are not physically associated
but happen to lie along the line of sight.
G6.55$-$0.10 appears to be elongated in the north-south direction 
in Figure 1$c$ (see also Fig. 1 of Andrews et al. 1985).
The central velocity of the \hrrl\ line emission from this source 
is 13.2~\kms, which is in good agreement with 
that ($\sim$15~\kms) of Andrews et al. (1985) 
who made high-resolution (8.$''$3$\times$4.$''$0) \hrrl\ line observations 
using the VLA.

\subsubsection{G8.14+0.23}
G8.14+0.23 is located at the peak of an extended ($4.'7 \times 3.'4$)
structure with a bipolar morphology 
extended in the north-south direction (Fig. 1$d$).  
Such a radio continuum appearance can be interpreted as a result of the
champagne flow originating from a thin, flat molecular cloud. 
The central velocity of the southern protuberance is
equal to that of the compact component within 2~\kms, which is possible 
if the axis of the extended HII region is in the plane of the sky,
similar to G10.30$-$0.15.
Since the surface brightness declines more rapidly on the eastern side
than on the western side, the gas density and/or the thickness
appear to decrease from the east to the west along the thin 
molecular cloud.
Shepherd \&\ Churchwell (1996) found high-velocity molecular gas,
which is suggestive of bipolar outflow, in this object.

\subsubsection{G12.21$-$0.10}
G12.21$-$0.10 is embedded in extended emission with 
three compact components (Fig. 1$g$).
The surface brightness declines slowly to the northeast, while it drops
steeply on the other side.
G12.21$-$0.10 appears as a $\sim$4$''$ cometary structure 
in the VLA map of WC89,
and is located in the northern compact component in our radio continuum map.
The compact components have the same velocity, which is different from 
that of the northeastern part by $\sim$3~\kms. 
High-velocity molecular gas was detected toward this object
(Shepherd \&\ Churchwell 1996).


\section{Conclusions}

We made VLA radio continuum observations of 16 simple \uchii\ regions 
with large ratios of $S_{\rm SD}$/$S_{\rm VLA}$ and detected 
extended emission towards all the sources.
The extended emission consists of one to several 
compact components and a diffuse envelope.
All the \uchii\ regions except for two spherical ones (G23.46$-$0.20 
and G25.72+0.05) are located in the compact components.
There is no significant velocity difference between
the ultracompact, compact, and extended components in each source
with one exception (G25.72+0.05), 
which strongly suggests that they are physically associated
in almost all of our sources. 
The \uchii\ regions always correspond to the peaks of their associated compact
components and
the compact components with \uchii\ regions are usually more compact than
those without \uchii\ regions. 
It is therefore likely that the associated compact components
and their immediate envelopes
have been produced by UV photons escaping from the \uchii\ regions,
although we can not completely exclude the possibility 
that they could have been created by separate ionizing sources 
in a cluster or by sequentially formed massive stars.
Consequently, 
almost all of the \uchii\ regions in our sample are unlikely to be 
`real' \uchii\ regions and their actual ages may be much greater than
10$^4$~yr.
The two \uchii\ regions with no associated compact components 
may be `real' \uchii\ regions or externally ionized dense clumps.

We calculated the ratios of $S_{\rm SD}$/$S_{\rm VLA}$ for
the previously known \uchii\ regions with simple morphology
and found that most of them have large ratios like our sources.
Therefore, 
the age problem of \uchii\ regions would be significantly alleviated.
We presented a simple model in which 
the presence of extended emission around \uchii\ regions
can be understood using both the Champagne flow model and
the hierarchical structure of massive star-forming regions.
In our model the \uchii\ regions and their associated compact and extended
components may not represent an evolutionary sequence. 
The three components could coexist for a long ($>$10$^5$~yr) time
for an \hii\ region which evolves 
within a hierarchically clumpy molecular cloud. 
In this context,
the compact components without \uchii\ regions seem to be in a later
evolutionary phase than those with \uchii\ regions.
The relation $n_{\rm e} \propto$~D$^{-1}$ may also be 
a corollary of the hierarchical structure of molecular clouds.

\acknowledgements
We thank Dana Balser for his help with the RRL observations
and data reduction. We are very grateful to Ed Churchwell and
Stan Kurtz for carefully reading the manuscript and
for many helpful comments and suggestions. 
K.-T. K. acknowledges support from Korea Research
Foundation grant for junior researcher in 1996.
This work has been supported by 
Korea Science and Engineering Foundation through Grant 
No. 961-0203-014-02, and also
BK21 Program, Ministry of Education, Korea through SEES.

\clearpage

\clearpage

\begin{deluxetable}{lc}
\footnotesize
\tablewidth{0pt}
\tablecaption{PARAMETERS OF VLA 21~CM CONTINUUM OBSERVATIONS}

\tablehead{
\colhead{Parameter} & \colhead{Value}
}
\startdata
Observation dates & 1995 Feb 11, 13, 14, $\&$ 17\\
Total time  & 19 hours \\
Configuration & DnC \\
Number of antennas & 27 \\
Center frequency  & 1.38510 $\&$ 1.46490 GHz\\
Bandwidth  & 50 MHz\\
Largest structure visible  & 15$'$ \\
Synthesized beam FWHM  & $\sim$40$''$ $\times$ 20$''$\\
Primary beam FWHM  & 30$'$\\
Flux calibrator source & 3C286 \\
Phase calibrator source & 1819-096\\
Typical on-source time & 40 minutes\\
Number of fields observed & 16 \\
Typical rms noise & 1 mJy~beam$^{-1}$\\
\enddata
\end{deluxetable}

\begin{deluxetable}{rccccccc}
\footnotesize
\tablewidth{0pt}
\tablecaption{OBSERVATIONAL PARAMETERS OF 16 \uchii\ REGIONS
FOR VLA 21~CM CONTINUUM OBSERVATIONS}
\tablehead{
  & \multicolumn{2}{c}{Phase} & \colhead{ON-source} & \colhead{Synthesized} & 
\colhead{RMS noise} & \colhead{Adopted\tablenotemark{a}} 
& \colhead{Morphology\tablenotemark{b}}\\
  & \multicolumn{2}{c}{Center} & \colhead{Time} & \colhead{FWHM} & 
\colhead{of Image}  & \colhead{Distance} & \colhead{of}\\ \cline{2-3}
\colhead{Source} & \colhead{$\alpha$(1950)} & \colhead{$\delta$(1950)} & 
\colhead{(minute)} & \colhead{($''$)} & \colhead{(mJy~beam$^{-1}$)} & 
\colhead{(kpc)} & \colhead{\uchii\ Region}
}
\startdata
~G5.89$-$0.39 & $17^{\rm h} 57^{\rm m} 26.\!\!^{\rm s}8$ & $-24^\circ 03' 56''$ & 40 & $39 \times 27$ & 1.70 & 
~2.6 & shell\\
~G5.97$-$1.17 & 18 ~00 ~36.4& $\!\!-24$ ~$\!$22 54 & 35 & $40 \times 28$ 
& 2.75 & ~1.9 & core-halo\\
~G6.55$-$0.10 & 17 ~57 ~47.4& $\!\!-23$ ~$\!$20 30 & 35 & $42 \times 25$ 
& 2.50 & 16.7 & irregular\\
~G8.14+0.23 & 18 ~00 ~00.2& $\!\!-21$ ~$\!$48 15 & 35 & $50 \times 22$ 
& 0.53 & ~4.2 & irregular\\
G10.15$-$0.34 & 18 ~06 ~22.5& $\!\!-20$ ~$\!$20 05 & 40 & $37 \times 25$ 
& 1.71 & ~6.0 & irregular\\
G10.30$-$0.15 & 18 ~05 ~57.9& $\!\!-20$ ~$\!$06 26 & 40 & $37 \times 25$ 
& 1.62 & ~6.0 & cometary\\
G12.21$-$0.10 & 18 ~09 ~43.7& $\!\!-18$ ~$\!$25 09 & 40 & $38 \times 24$ 
& 0.49 & 16.1 & cometary\\
G12.43$-$0.05 & 18 ~09 ~58.9& $\!\!-18$ ~$\!$11 58 & 40 & $40 \times 24$ 
& 0.90 & 16.7 & cometary\\
G23.46$-$0.20 & 18 ~32 ~01.2& $\!\!-08$ ~$\!$33 33 & 45 & $40 \times 18$ 
& 0.91 & ~9.0 & spherical\\
G23.71+0.17 & 18 ~31 ~10.3& $\!\!-08$ ~$\!$09 36 & 45 & $40 \times 18$ 
& 0.73 & ~9.0 & core-halo\\
G23.96+0.15 & 18 ~31 ~42.6& $\!\!-07$ ~$\!$57 11 & 60 & $40 \times 18$ 
& 0.60 & ~6.0 & irregular\\
G25.72+0.05 & 18 ~35 ~21.6& $\!\!-06$ ~$\!$26 27 & 55 & $41 \times 18$ 
& 0.85 & 14.0 & spherical\\
G27.28+0.15 & 18 ~37 ~55.7& $\!\!-05$ ~$\!$00 35 & 60 & $40 \times 17$ 
& 0.65 & 15.2 & irregular\\
G29.96$-$0.02 & 18 ~43 ~27.1& $\!\!-02$ ~$\!$42 36 & 50 & $44 \times 18$ 
& 1.62 & ~9.0 & cometary\\
G35.05$-$0.52 & 18 ~54 ~37.1& $\!\!+01$ ~$\!$35 01 & 35 & $44 \times 17$ 
& 1.37 & 12.7 & irregular\\
G37.55$-$0.11 & 18 ~57 ~46.8& $\!\!+03$ ~$\!$58 55 & 35 & $44 \times 17$ 
& 0.61 & 12.0 & core-halo 
\enddata
\tablenotetext{a}{From references given by Wood $\&$ Churchwell (1989)}
\tablenotetext{b}{Wood $\&$ Churchwell (1989)}
\end{deluxetable}

\clearpage
 
\begin{deluxetable}{rcccccc}
\footnotesize
\tablewidth{0pt}
\tablecaption{21~CM CONTINUUM PARAMETERS OF 16 \uchii\ REGIONS}
\tablehead{
  & \multicolumn{2}{c}{Peak} & \multicolumn{2}{c}{Extent\tablenotemark{a}}
& \colhead{Peak} & \colhead{Integrated} \\ \cline{4-5}
  & \multicolumn{2}{c}{Position} & \colhead{Angular} & \colhead{Linear}
& \colhead{Brightness} & \colhead{Flux Density} \\ \cline{2-3}
\colhead{Source} & \colhead{$\alpha$(1950)} & \colhead{$\delta$(1950)} &
\colhead{(square arcmin)} & \colhead{(pc$^2$)} & \colhead{(Jy~beam$^{-1}$)} &
\colhead{(Jy)} 
}
\startdata
~G5.89$-$0.39 & $17^{\rm h} 57^{\rm m} 37.\!\!^{\rm s}7$& $-24^\circ 03' 51''$&
$14.4 \times ~9.0$ & $10.9 \times ~6.8$ & 0.91 & 20.86 \\
A ............ & 17 ~57 ~37.7& $\!\!-24$ ~$\!$04 08 &
$1.58 \times 1.38$ & $1.20 \times 1.05$ & 0.91 & ~7.79 \\
B\tablenotemark{\dagger} ............ & 17 ~57 ~27.3& $\!\!-24$ ~$\!$04 13 & 
$1.09 \times 0.79$ & $0.82 \times 0.60$ & 0.32 & ~1.30 \\
~G5.97$-$1.17 & 18 ~00 ~36.8& $\!\!-24$ ~$\!$22 49 & $14.5 \times 10.7$ &
$~8.0 \times ~5.9$ & 3.57 &48.29 \\
A\tablenotemark{\dagger}  ............ & 18 ~00 ~36.6& $\!\!-24$ ~$\!$22 48 &
$1.24 \times 1.14$ & $0.69 \times 0.63$ & 2.69 & 14.93 \\
~G6.55$-$0.10 & 17 ~57 ~47.0& $\!\!-23$ ~$\!$20 25 & $~3.5 \times ~2.2$ &
$17.0 \times 10.7$ & 1.07 & ~2.32 \\
A\tablenotemark{\dagger}  ............ & 17 ~57 ~47.0& $\!\!-23$ ~$\!$20 24 &
$0.55 \times 0.48$ & $2.67 \times 2.33$ & 1.03 & ~2.06 \\
~G8.14+0.23 & 18 ~00 ~00.9& $\!\!-21$ ~$\!$48 15 & $~6.9 \times ~2.8$ &
$~8.4 \times ~3.4$ & 2.04&~6.67 \\
A\tablenotemark{\dagger}  ............ & 18 ~00 ~00.5& $\!\!-21$ ~$\!$48 13 &
$0.75 \times 0.57$ & $0.92 \times 0.70$ & 1.74 & ~4.71 \\
G10.15$-$0.34 & 18 ~06 ~27.8& $\!\!-20$ ~$\!$20 00 & $10.9 \times ~6.7$ &
$19.0 \times 11.7$ & 2.94& 55.22  \\
A  ............ & 18 ~06 ~27.5& $\!\!-20$ ~$\!$19 59 &
$1.75 \times 1.08$ & $3.05 \times 1.88$ & 2.77 & 23.20 \\
B\tablenotemark{\dagger} ............ & 18 ~06 ~22.0& $\!\!-20$ ~$\!$20 00 &
$1.57 \times 0.48$ & $2.74 \times 0.84$ & 2.24 & ~9.94 \\
G10.30$-$0.15 & 18 ~06 ~02.1& $\!\!-20$ ~$\!$05 41 & $12.8 \times ~4.6$ &
$22.3 \times ~8.0$ & 1.42& 15.92  \\
A ............  & 18 ~06 ~02.5& $\!\!-20$ ~$\!$05 31 &
$1.28 \times 0.76$ & $2.23 \times 1.33$ & 1.32 & ~6.63 \\
B\tablenotemark{\dagger} ............  & 18 ~05 ~57.8& $\!\!-20$ ~$\!$06 19 &
$0.93 \times 0.63$ & $1.62 \times 1.10 $ & 1.03 & ~3.53 \\
G12.21$-$0.10 & 18 ~09 ~42.0& $\!\!-18$ ~$\!$26 05 & $~4.7 \times ~3.4$ &
$22.0 \times 15.9$ & 0.32& ~3.06 \\
A ............ & 18 ~09 ~42.1& $\!\!-18$ ~$\!$25 59 &
$0.56 \times 0.47$ & $2.65 \times 2.19$ & 0.32 & ~0.67 \\
B ............ & 18 ~09 ~47.7& $\!\!-18$ ~$\!$26 33 &
$1.25 \times 0.82$ & $5.85 \times 3.82$ & 0.24 & ~1.27 \\
C\tablenotemark{\dagger} ............ & 18 ~09 ~44.0& $\!\!-18$ ~$\!$25 15 &
$0.85 \times 0.46$ & $3.98 \times 2.19$ & 0.18 & ~0.46 \\
G12.43$-$0.05 & 18 ~09 ~58.9& $\!\!-18$ ~$\!$11 58 & $~3.0 \times ~2.2$ &
$14.6 \times 10.7$ & 0.14& ~0.69 \\
A\tablenotemark{\dagger} ............ & 18 ~09 ~58.7& $\!\!-18$ ~$\!$11 56 &
$1.02 \times 0.78$ & $4.95 \times 3.79$ & 0.11 & ~0.45 \\
G23.46$-$0.20 & 18 ~32 ~00.5& $\!\!-08$ ~$\!$34 43 & $~8.8 \times ~5.8$ &
$23.0 \times 15.2$ & 0.41& 11.31 \\
A ............ & 18 ~32 ~00.7& $\!\!-08$ ~$\!$34 48 &
$1.98 \times 0.88$ & $5.18 \times 2.30$ & 0.37 & ~3.47 \\
B ............ & 18 ~32 ~01.1& $\!\!-08$ ~$\!$36 10 & 
$3.47 \times 1.35$ & $9.08 \times 3.53$ & 0.28 & ~6.80 \\
G23.71+0.17 & 18 ~31 ~10.0& $\!\!-08$ ~$\!$09 41 & $~2.9 \times ~1.8$ &
$~7.6 \times ~4.7$ & 0.46& ~1.93 \\
A\tablenotemark{\dagger} ............ & 18 ~31 ~10.0& $\!\!-08$ ~$\!$09 50 & 
$0.95 \times 0.57$ & $2.49 \times 1.49$ & 0.44 & ~1.85 \\
G23.96+0.15 & 18 ~31 ~42.2& $\!\!-07$ ~$\!$57 11 & $~3.1 \times ~1.5$ &
$~5.4 \times ~2.6$ & 1.08 & ~2.13 \\
A\tablenotemark{\dagger} ............ & 18 ~31 ~42.1& $\!\!-07$ ~$\!$57 10 &
$0.43 \times 0.25$ & $0.75 \times 0.44$ & 1.02 & ~1.71 \\
G25.72+0.05 & 18 ~35 ~22.9& $\!\!-06$ ~$\!$28 12 & $~3.2 \times ~2.5$ &
$13.0 \times 10.2$ & 0.18 & ~1.74  \\
A ............ & 18 ~35 ~22.3& $\!\!-06$ ~$\!$28 04 & 
$1.43 \times 1.11$ & $5.82 \times 4.52$ & 0.16 & ~1.43 \\
G27.28+0.15 & 18 ~37 ~55.7& $\!\!-05$ ~$\!$00 35 & $~2.6 \times ~1.4$ &
$11.5 \times ~6.2$ & 0.44& ~0.96  \\
A\tablenotemark{\dagger} ............ & 18 ~37 ~55.7& $\!\!-05$ ~$\!$00 35 &
$0.57 \times 0.29$ & $2.52 \times 1.28$ & 0.40 & ~0.74 \\
G29.96$-$0.02 & 18 ~43 ~27.4& $\!\!-02$ ~$\!$42 36 & $~6.3 \times ~5.2$ &
$16.5 \times 13.6$ & 1.50& 12.69 \\
A\tablenotemark{\dagger} ............ & 18 ~43 ~27.4 & $\!\!-02$ ~$\!$42 34 & 
$0.40 \times 0.37$ & $1.05 \times 0.97$ & 1.36 & ~2.61 \\
B ............ & 18 ~43 ~32.6& $\!\!-02$ ~$\!$44 57 & 
$1.47 \times 1.23$ & $3.85 \times 3.22$ & 0.61 & ~6.08 \\
G35.05$-$0.52 & 18 ~54 ~31.8& $\!\!+01$ ~$\!$34 26 & $~3.7 \times ~3.1$ &
$13.7 \times 11.5$ & 0.19& ~1.64 \\
A\tablenotemark{\dagger} ............  & 18 ~54 ~37.0& $\!\!+01$ ~$\!$34 56 & 
$0.43 \times 0.27$ & $1.59 \times 1.00$ & 0.17 & ~0.27 \\
B ............ & 18 ~54 ~31.9& $\!\!+01$ ~$\!$34 44 & 
$1.56 \times 1.02$ & $5.76 \times 3.77$ & 0.14 & ~1.37 \\
G37.55$-$0.11 & 18 ~57 ~46.8& $\!\!+03$ ~$\!$58 55 & $~2.1 \times ~1.8$ &
$~7.3 \times ~6.3$ & 0.45&~1.12 \\
A\tablenotemark{\dagger} ............  & 18 ~57 ~46.8& $\!\!+03$ ~$\!$58 53 & 
$0.54 \times 0.46$ & $1.88 \times 1.61$ & 0.39 & ~0.96 \\ 
\enddata
\tablenotetext{\dagger}{Compact component that contains an \uchii\ region}
\tablenotetext{a}{Deconvolved diameters at half-maximum for compact 
components, while diameters at 10~mJy~beam$^{-1}$ for extended envelopes}
\end{deluxetable}

\clearpage

\begin{deluxetable}{rccccclcc}
\footnotesize
\tablewidth{0pt}
\tablecaption{PHYSICAL PARAMETERS OF 16 \uchii\ REGIONS}
\tablehead{
  & \colhead{\nee} & \colhead{$EM$} & \colhead{$U$} & \colhead{\mhii} &
\colhead{log~\nc} & \colhead{Spectral\tablenotemark{a}} & 
\colhead{$f_{d}$} & \colhead{\te\tablenotemark{b}}\\
\colhead{Source} & \colhead{(cm$^{-3}$)} &
\colhead{(10$^4$ pc~cm$^{-6}$)} & \colhead{(pc~cm$^{-2}$)} & \colhead{(\msol)}
&  \colhead{(s$^{-1}$)} & \colhead{Type} & \colhead{(\%)} & \colhead{(K)}
}
\startdata
~G5.89$-$0.39 & ~~64 & ~~3.5 & ~69.1 & ~480 & 49.08 & O6 (O7) & 95& ~8500\\
   A ............ & ~473 & ~36.9 &  ~50.0 & ~~25 & 48.66 & O7 &     & \\
   B\tablenotemark{\dagger} ............ & ~391 & ~15.7 & ~27.4 & ~~~5 & 47.88 & O9.5 & & \\
~G5.97$-$1.17 & ~~98 & ~~6.6 & ~73.3 & ~380  & 49.20 & O6 (B0) & 
            $\!\!\sim$0 &~7700\\
   A\tablenotemark{\dagger} ............ & 1036 & 103.9 & ~49.6 & ~~11 & 48.69 & O7 & & \\
~G6.55$-$0.10 & ~~66 & ~~6.0 & 111.7 & 1970 & 49.80 & O4.5 (O8.5) & 34 & ~6700\\
   A\tablenotemark{\dagger} ............ & ~443 & ~72.5 & 107.3 & ~262 & 49.75 & O4.5 & & \\
~G8.14+0.23   & ~111 & ~~6.6 & ~62.0 & ~200 & 49.10 & O6 (B0) & 
            $\!\!\sim$0 &~5600\\
   A\tablenotemark{\dagger} ............ & ~917& ~98.3 & ~55.2 & ~~17 & 48.95 & O6.5 &  & \\
G10.15$-$0.34 & ~~99 & ~14.5 & 158.7 & 3800 & 50.33 & $>$O4 (O8) & 59 &~5500\\
   A ............ & ~555 & 108.3 & 118.9 & ~285 & 49.95 & O4 &   & \\
   B\tablenotemark{\dagger} ............ & ~726 & 116.8 & ~89.6 & ~~93 & 49.58 & O5 && \\
G10.30$-$0.15 & ~~64 & ~~5.5 & 107.5 & 1820 & 49.74 & O4.5 (O8.5) & 73 &~6800\\
   A ............ & ~505 & ~64.5 & ~80.2 & ~~96 & 49.36 & O5.5 &   & \\
   B\tablenotemark{\dagger} ............ & ~540 & ~57.2 & ~65.0 & ~~48 & 49.09 & O6  && \\
G12.21$-$0.10 & ~~46 & ~~3.9 & 120.2 & 3550 & 49.88 & O4 (O6.5) & 88 & ~7000\\
   A ............ & ~261 & ~24.2 & ~72.4 & ~137 & 49.22 & O5.5    &  & \\
   B ............ & ~136 & ~11.9 & ~89.6 & ~518 & 49.50 & O5    &  & \\
   C\tablenotemark{\dagger} ............ & ~160 & ~11.0 & ~63.9 & ~153 & 49.05 & O6 && \\
G12.43$-$0.05 & ~~44 & ~~2.4 & ~78.2 & 1030 & 49.19 & O6 (O9) & 
             $\!\!\sim$0 &10000\\
   A\tablenotemark{\dagger} ............ & ~~99 & ~~6.2 & ~68.0 & ~297 & 49.00 & O6 && \\
G23.46$-$0.20 & ~~49 & ~~4.5 & 124.6 & 3700 & 49.96 & O4 (B0) & 49 &~6300\\
   A ............ & ~191 & ~18.4 & ~84.0 & ~292 & 49.45 & O5.5 &   & \\
   B ............ & ~126 & ~13.3 & 105.1 & ~864 & 49.74 & O5   && \\
G23.71+0.17   & ~112 & ~~7.6 & ~70.1 & ~290 & 49.17 & O6 (O9) & 38 & ~7100\\
   A\tablenotemark{\dagger} ............ & ~341 & ~32.8 & ~69.0 & ~~91 & 49.15 & O6 && \\
G23.96+0.15   & ~145 & ~~8.1 & ~53.1 & ~100 & 48.93 & O6.5 (O9.5) & 42 & ~5100\\
   A\tablenotemark{\dagger} ............ & 1299 & 140.0 & ~49.4 & ~~~9 & 48.84 & O6.5 && \\
G25.72+0.05   & ~~63 & ~~4.5 & ~90.2 & 1100 & 49.52 & O5 (B0) & 35 & ~6700\\
   A ............ & ~106 & ~~8.5 & ~84.5 & ~534 & 49.43 & O5.5  &  & \\
G27.28+0.15   & ~~82 & ~~5.7 & ~79.5 & ~570 & 49.30 & O5.5 (O9.5)& 48 & ~7700\\
   A\tablenotemark{\dagger} ............ & ~418 & ~45.5 & ~72.9 & ~~87 & 49.19 & O6 && \\
G29.96$-$0.02 & ~~71 & ~~7.5 & 128.2 & 2780 & 50.03 & O4 (O5.5) & 85 &~5800\\
  A\tablenotemark{\dagger} ............ & 1022 & 155.7 & ~75.7 & ~~40 & 49.34 & O5.5  & & \\
  B ............ & ~241 & ~30.1 & 100.3 & ~393 & 49.71 & O5    &  & \\
G35.05$-$0.52 & ~~50 & ~~3.1 & ~84.8 & 1160 & 49.37 & O5.5 (O9) & 
          83 & ~8100\\
   A\tablenotemark{\dagger} ............ & ~356 & ~23.5 & ~46.7 & ~~27 & 48.59 & O7 & &\\
   B ............ & ~112 & ~~8.7 & ~79.9 & ~426 & 49.29 & O5.5 &  & \\
G37.55$-$0.11 & ~~96 & ~~6.1 & ~69.5 & ~330 & 49.21 & O6 (O7.5) & 52 &~6100\\
   A\tablenotemark{\dagger} ............ & ~376 & ~35.9 & ~66.0 & ~~72 & 49.15 & O6 & &\\
\enddata
\tablenotetext{\dagger}{Compact component that contains an \uchii\ region}
\tablenotetext{a}{Spectral types measured from \uchii\ regions alone 
are given in parentheses} 
\tablenotetext{b}{Taken from Downes et al. (1980) and Wink et al.
(1982)}
\end{deluxetable}

\clearpage

\begin{deluxetable}{rccccc}
\footnotesize
\tablewidth{0pt}
\tablecaption{H76$\alpha$ LINE PARAMETERS OF \uchii\ REGIONS}
\tablehead{
  & \colhead{Observed Position} & \colhead{\vlsr} & 
\colhead{\tl} & \colhead{\delv} & \colhead{$\int T_{\rm L} dv$} \\
\cline{2-2}
\colhead{Source} & \colhead{$\alpha$(1950) ~~~~~~~$\delta$(1950)} & 
\colhead{(\kms)} & 
\colhead{(mK)} & \colhead{(\kms)} & \colhead{(mK~\kms)}
}
\startdata
~G5.89$-$0.39\tablenotemark{a} & 
~$17^{\rm h} 57^{\rm m} 26.\!\!^{\rm s}8$ ~~$-24^\circ 03' 56''$ 
&  &  &  &   \\
  & (~~$00.\!\!^{\rm s}0$, ~~~$\!$$00' 00''$)\tablenotemark{\circ}  & 
~~9.9$\pm$0.4 & ~95$\pm$1.4 &  54.8$\pm$1.0& 5484   \\
  & ($-$13.4, ~~00 00) & ~15.0$\pm$0.8 & ~25$\pm$1.3 &  
30.7$\pm$1.9 & ~771  \\
  & ($-$05.4, ~~00 00) & ~13.5$\pm$0.7 & ~29$\pm$1.7 &  
22.9$\pm$1.5 & ~684   \\
  & (+02.6, ~~00 00)\tablenotemark{\circ} & 
~~4.6$\pm$0.5 & ~95$\pm$1.7  &  53.5$\pm$1.1 & 5436   \\
  & (+10.6, $-$04 00) & ~15.0$\pm$0.5  & ~34$\pm$1.7   &  
19.7$\pm$1.1 & ~757    \\
  & (+10.6, $-$02 00) & ~12.3$\pm$0.2  & ~56$\pm$1.2  &  
21.1$\pm$0.5 & 1253  \\
  & (+10.6, ~~00 00)\tablenotemark{\bullet} & 
~10.4$\pm$0.1  & 144$\pm$1.5  &  24.7$\pm$0.3 & 3955   \\
  & (+10.6, +02 00) & ~11.7$\pm$0.3  & ~55$\pm$1.5  &  
23.3$\pm$0.7& 1404  \\
  & (+10.6, +04 00) & ~10.7$\pm$0.7 & ~26$\pm$1.5  &  
26.1$\pm$1.7 & ~709   \\
  & (+18.6, ~~00 00)\tablenotemark{\circ} & 
~~9.8$\pm$0.2  & 122$\pm$1.8  & 26.7$\pm$0.4 & 3594   \\
  & (+26.6, ~~00 00) & ~13.5$\pm$0.5  & ~43$\pm$1.7  &  
23.8$\pm$1.1 & 1083   \\
  & (+34.6, ~~00 00) & ~~8.1$\pm$0.6  & ~33$\pm$1.9  &  
20.9$\pm$1.4 & ~739   \\
  & (+42.6, ~~00 00) &  & $\leq7$ &  &   \\
%
%
~G5.97$-$1.17\tablenotemark{a} & 18 ~$\!$00 ~36.4  ~~$-24$ ~$\!$22 54 
&  & & &   \\
  & (~~00.0, ~~~$\!$00 00)\tablenotemark{\bullet} & 
~~4.5$\pm$0.1 & 279$\pm$1.7 & 26.8$\pm$0.2 & 7856    \\
  & ($-$24.0, ~~00 00)  & ~~3.3$\pm$0.4  & ~37$\pm$1.6  &  
20.2$\pm$1.0 & ~789   \\
  & ($-$16.0, ~~00 00)  & ~~2.4$\pm$0.3  & ~67$\pm$1.7  &  
22.5$\pm$0.7 & 1568   \\
  & ($-$08.0, ~~00 00)  & ~~2.5$\pm$0.2  & 120$\pm$1.8  &  
23.1$\pm$0.4 & 2976   \\
  & (~~00.0, $-$06 00)  & ~~5.0$\pm$0.5  & ~40$\pm$2.0  &  
19.7$\pm$1.1 & ~874  \\
  & (~~00.0, $-$04 00)  & ~~4.5$\pm$0.2 & ~61$\pm$1.3  &  
20.0$\pm$0.5 & 1236   \\
  & (~~00.0, $-$02 00)\tablenotemark{\circ} & 
~~3.8$\pm$0.2  & 120$\pm$1.8  &  24.7$\pm$0.4 & 3183   \\
  & (~~00.0, +02 00)\tablenotemark{\circ}  & 
~~2.3$\pm$0.1  & 185$\pm$1.9  &  25.0$\pm$0.3 & 4906   \\
  & (~~00.0, +04 00)  & ~~1.5$\pm$0.2  & ~80$\pm$1.8  &  
22.0$\pm$0.6 & 1827   \\
  & (~~00.0, +06 00)  & ~$\!\!\!-$0.2$\pm$0.5  & ~46$\pm$1.9  &  
24.1$\pm$1.2 & 1244   \\
  & (+08.0, ~~00 00)  & ~~4.6$\pm$0.1  & 129$\pm$1.3  &  
25.2$\pm$0.3 & 3420   \\
  & (+16.0, ~~00 00)  & ~~6.9$\pm$0.2  & ~91$\pm$1.6  &  
21.1$\pm$0.4 & 2140   \\
  & (+24.0, ~~00 00)  & ~~7.8$\pm$0.3  & ~62$\pm$1.8  &  
22.7$\pm$0.8 & 1492   \\
  & (+32.0, ~~00 00)  & ~~5.0$\pm$0.7  & ~40$\pm$2.4  &  
22.2$\pm$1.5 & ~865   \\
  & (+40.0, ~~00 00)  & ~~3.0$\pm$0.4  & ~60$\pm$2.0  &  
21.8$\pm$0.8 & 1456   \\
  & (+08.0, $-$04 00)  & ~~3.4$\pm$0.3 & ~60$\pm$1.9  &  
18.9$\pm$0.7 & 1182   \\
  & (+16.0, $-$04 00)  & ~~1.6$\pm$0.2  & ~85$\pm$1.4  &  
22.1$\pm$0.4 & 2002   \\
  & (+24.0, $-$04 00)  & ~~0.5$\pm$0.3 & ~80$\pm$2.1  &  
19.9$\pm$0.6 & 1669   \\
  & (+32.0, $-$04 00)  & ~~1.8$\pm$0.5 & ~46$\pm$1.8  &  
30.0$\pm$1.2 & 1341   \\
  & (+40.0, $-$04 00)  & ~~1.3$\pm$0.5 & ~44$\pm$2.0  &  
21.8$\pm$1.2 & ~976   \\
%
%
~G6.55$-$0.10 & 17 ~$\!$57 ~47.4  ~~$-23$ ~$\!$20 30 & &  & &   \\
  & (~~00.0, ~~~$\!$00 00)  & 
~13.2$\pm$0.4 & ~70$\pm$1.9 & 26.5$\pm$0.8 & 2025   \\
%
%
~G8.14+0.23  & 18 ~$\!$00 ~00.2  ~~$-21$ ~$\!$48 15 & & & &  \\
  & (~~00.0, ~~~$\!$00 00)  & 
~20.3$\pm$0.4 & ~58$\pm$1.6 & 27.3$\pm$0.9 & 1700  \\
  & (+06.4, $-$01 10) & ~22.2$\pm$1.1 & ~19$\pm$1.8 & 
22.2$\pm$2.5 & ~405    \\
\hline
\tablebreak
%
%
G10.15$-$0.34 & 18 ~$\!$06 ~22.5  ~~$-20$ ~$\!$20 05 & & &  &  \\
  & (~~00.0, ~~~$\!$00 00)\tablenotemark{\bullet}  &
~16.4$\pm$0.2 & 163$\pm$2.3 & 29.7$\pm$0.5 & 5170  \\
  & ($-$10.8, ~~00 00) & ~23.5$\pm$0.9 & ~48$\pm$1.9 &
47.5$\pm$2.1 & 2193   \\
  & ($-$05.8, +02 00)\tablenotemark{\circ} & ~14.6$\pm$0.2 &
190$\pm$2.0 & 35.8$\pm$0.5 & 7473   \\
  & ($-$02.8, $-$02 00) & ~13.5$\pm$0.9 & ~18$\pm$0.2 &
16.5$\pm$2.0 & ~344   \\
  & ($-$02.8, ~~00 00) & ~16.2$\pm$0.1 & 254$\pm$2.1  &
32.5$\pm$0.3 & 8819   \\
  & (~~00.0, +05 00) & ~14.3$\pm$0.6 & ~30$\pm$1.4 &
26.1$\pm$1.4 & ~859   \\
  & (+03.8, +02 25)\tablenotemark{\bullet} &
~15.4$\pm$0.2 & 306$\pm$3.1 & 32.1$\pm$0.4 & $\!\!\!$10598   \nl
  & (+05.2, $-$04 00) &    & $\leq$8 &   &   \\
  & (+05.2, $-$02 00) & ~11.0$\pm$0.5 & ~47$\pm$1.6 &
31.3$\pm$1.2 & 1517   \\
  & (+05.2, +00 10)\tablenotemark{\bullet} &
~14.6$\pm$0.1 & 352$\pm$2.4 & 32.0$\pm$0.4 & $\!\!\!$12226   \nl
  & (+06.3, $-$05 40) & ~15.0$\pm$0.3 & ~40$\pm$1.3 &
19.6$\pm$0.7 & ~806   \nl
  & (+13.2, $-$02 00) & ~18.4$\pm$0.6 & ~55$\pm$1.5 &
42.9$\pm$1.3 & 2419   \\
  & (+13.8, ~~00 00) & ~10.3$\pm$0.1 & 189$\pm$1.2 &
32.7$\pm$0.3 & 6712   \\
  & (+22.7, $-$00 10) & ~~4.5$\pm$0.4 & ~37$\pm$1.4 &
22.6$\pm$1.0 & ~840   \\
  & (+30.4, ~~00 00) & ~~4.0$\pm$0.9 & ~21$\pm$1.8 &
19.9$\pm$2.1 & ~567  \\
%
%
G10.30$-$0.15  & 18 ~$\!$05 ~57.9  ~~$-20$ ~$\!$06 26 & & & &   \\
  & (~~00.0, ~~~$\!$00 00)  & 
~~7.7$\pm$0.5 & ~49$\pm$1.6 & 28.5$\pm$1.1 & 1479  \\
  & ($-$06.8, +02 10) & ~11.9$\pm$0.4 & ~31$\pm$1.2 & 
23.0$\pm$1.0 & ~833   \\
  & (+03.5, $-$01 55) & ~~8.1$\pm$0.5 & ~38$\pm$1.3 & 
25.7$\pm$1.1 & 1006    \\
  & (+04.2, +00 50)\tablenotemark{\bullet} & 
~11.1$\pm$0.2 & 122$\pm$1.9 & 29.7$\pm$0.5 & 3772   \\
%
%
G12.21$-$0.10  & 18 ~$\!$09 ~43.7  ~~$-18$ ~$\!$25 09 &  &  & &  \\
  & ($-$01.0, $-$00 20)\tablenotemark{b} & ~28.9$\pm$0.4 & 
~31$\pm$1.1 & 22.2$\pm$0.9 & ~787  \\
  & (+04.2, $-$01 25) & ~27.8$\pm$0.4 & ~40$\pm$1.7 & 
19.5$\pm$0.9 & ~960   \\
  & (+05.6, +00 45) & ~25.3$\pm$0.4 & ~32$\pm$1.7 & 
17.0$\pm$1.0 & ~545    \\
%
%
G12.43$-$0.05  & 18 ~$\!$09 ~58.9  ~~$-18$ ~$\!$11 58 &  & & &  \\
  & (~~00.0, ~~~$\!$00 00) & 
 & $\leq$7 & &  \\
%
%
G23.46$-$0.20  & 18 ~$\!$32 ~01.2  ~~$-08$ ~$\!$33 33 & & & &   \\
  & (~~00.0, ~~~$\!$00 00) & 
~99.0$\pm$0.3 & 54$\pm$2.0 & 16.7$\pm$0.8 & ~912  \\
  & ($-$10.8, $-$02 00) & ~99.9$\pm$0.3 & ~52$\pm$1.5 & 
21.5$\pm$0.7 & 1166   \\
  & ($-$04.7, +00 30) & ~98.7$\pm$0.3 & ~47$\pm$1.7 & 
16.3$\pm$0.7 & ~852  \\
  & ($-$01.3, $-$03 00) & 103.7$\pm$0.3 & ~54$\pm$1.4 & 
21.8$\pm$0.7 & 1224   \\
  & ($-$00.8, $-$02 00)\tablenotemark{\bullet} & 
103.1$\pm$0.2 & 126$\pm$1.6 &  24.4$\pm$0.4 & 3293  \\
  & (~~00.0, $-$01 10) & 
103.3$\pm$0.2 & 103$\pm$0.2 & 23.0$\pm$0.5 & 2508  \\
  & (+09.1, $-$02 00) & 105.2$\pm$1.8 & ~48$\pm$1.8 & 
19.7$\pm$0.9 & ~984   \\
  & (+27.0, $-$02 05) & & $\leq$6 &  &   \\
%
%
G23.71+0.17  & 18 ~$\!$31 ~10.3  ~~$-08$ ~$\!$09 36 & & & & \\
  & (~~00.0, ~~~$\!$00 00) & 
103.0$\pm$0.6 & ~39$\pm$1.7 & 30.3$\pm$1.5 & 1292  \\
%
%
G23.96+0.15  & 18 ~$\!$31 ~42.6  ~~$-$07 ~$\!$57 11 & & & &  \\
  & (~~00.0, ~~~$\!$00 00) & 
~78.9$\pm$0.5 & ~49$\pm$1.7 & 28.0$\pm$1.2 & 1295  \\
%
%
G25.72+0.05  &  18 ~$\!$35 ~21.6, ~~$-06$ ~$\!$26 27 & & &   \\
  & (~~00.0, ~~~$\!$00 00) & 
~98.3$\pm$1.9 & 12$\pm$1.5 & 30.9$\pm$4.6 & 405  \\
  & (+01.3, $-$01 45) & 
~54.1$\pm$0.4 & ~32$\pm$1.0 & 27.8$\pm$1.0 & ~948  \\
\hline
\tablebreak
%
%
G27.28+0.15  & 18 ~$\!$37 55.7  ~~$-05$ ~$\!$00 35  & & & &  \\
  & (~~00.0, ~~~$\!$00 00) & 
~35.0$\pm$0.8 & ~20$\pm$1.3 & 23.9$\pm$1.9 & ~472 \\
%
%
G29.96$-$0.02 & 18 ~$\!$43 ~27.1  ~~$-02$ ~$\!$42 36 & & & &  \\
  & (~~00.0, ~~~$\!$00 00)\tablenotemark{\bullet} & 
~95.9$\pm$0.2 & 115$\pm$1.7 & 28.6$\pm$0.6 & 3730  \\
  & ($-$08.0, $-$03 05) & ~98.9$\pm$0.2 & ~68$\pm$1.4 & 
22.7$\pm$0.6 & 1667  \\
  & ($-$01.7, $-$06 15) & ~98.0$\pm$0.3 & ~44$\pm$1.1 & 
23.7$\pm$0.7 & 1122   \\
  & ($-$00.6, $-$04 45) & 
~99.1$\pm$0.4 & ~54$\pm$1.7 &  24.4$\pm$1.0 & 1388   \\
  & ($-$00.6, $-$03 15)\tablenotemark{\bullet} & 
100.0$\pm$0.3 & 121$\pm$2.5 & 26.4$\pm$0.7 & 3714   \\
  & ($-$00.6, $-$01 37)\tablenotemark{\circ} & 
~98.8$\pm$0.2  & ~81$\pm$1.3 &  25.6$\pm$0.5 & 2196   \\
  & (+05.6, $-$01 55)\tablenotemark{\bullet} & 
100.4$\pm$0.2 & 133$\pm$1.8 & 26.3$\pm$0.4 & 3646   \\
  & (+07.6, $-$03 10)\tablenotemark{\circ} & 
101.4$\pm$0.2 & 114$\pm$1.6 &  27.5$\pm$0.5 & 3346    \\
%
%
G35.05$-$0.52 & 18 ~$\!$54 ~37.1  ~~$+01$ ~$\!$35 01 & & & &  \\
  & (~~00.0, ~~~$\!$00 00) & 
~51.2$\pm$0.4 & ~26$\pm$1.1 & 17.9$\pm$0.9 & ~470  \\
  & ($-$08.0, +00 55) & ~50.0$\pm$0.7 & ~17$\pm$1.5 & 
16.9$\pm$1.7 & ~339  \\
  & ($-$05.7, $-$00 25) & ~52.0$\pm$0.4 & ~21$\pm$0.1 & 
17.9$\pm$1.0 & ~388  \\
  & ($-$02.9, $-$00 13) & ~52.7$\pm$0.4 & ~31$\pm$1.2 & 
~18.9$\pm$0.8 & ~600  \\
%
%
G37.55$-$0.11 & 18 ~$\!$57 ~46.8, ~~$+03$ ~$\!$58 55 & & & &  \\ 
  & (~~00.0, ~~~$\!$00 00) & 
~51.5$\pm$0.5 & ~28$\pm$0.1 & 30.0$\pm$1.3 & ~858  \\
\enddata
\tablenotetext{\circ}{Position where He76$\alpha$ line emission 
was observed but not detected (\tl$\lesssim$6~mK)}
\tablenotetext{\bullet}{Position where He76$\alpha$ line emission 
was detected (see Table 6)}
\tablenotetext{a}{Mapped sources for which part of spectra were presented}
\tablenotetext{b}{\uchii\ region is located within the beam}
\end{deluxetable}

\clearpage

\begin{deluxetable}{rcccccc}
\footnotesize
\tablewidth{0pt}
\tablecaption{He76$\alpha$ LINE PARAMETERS OF \uchii\ REGIONS}
\tablehead{
  &  & \colhead{\vlsr} &
\colhead{\tl} & \colhead{\delv} & \colhead{$\int T_{\rm L} dv$} & \\
\colhead{Source} & \colhead{Offsets\tablenotemark{a}} &
\colhead{(\kms)} &
\colhead{(mK)} & \colhead{(\kms)} & \colhead{(mK~\kms)} & \colhead{Y$^+$}
}
\startdata
~G5.89$-$0.39 & (+$10.\!\!^{\rm s}6$, ~~$00' 00''$) & ~~9.8$\pm$0.7 & 
11$\pm$1.7 & ~8.5$\pm$1.6 &  ~81 & 0.020 \\
~G5.97$-$1.17 & (~~00.0, ~~~$\!$00 00) & ~~1.5$\pm$0.6 &
21$\pm$2.1 &  12.1$\pm$1.4 & 275 & 0.035 \\
G10.15$-$0.34 & (~~00.0, ~~~$\!$00 00) & ~12.3$\pm$1.5 & 
12$\pm$2.8 & 13.2$\pm$3.5 & 179 & 0.035 \\
  & (+03.8, +02 25) & ~14.5$\pm$1.4 & 11$\pm$1.9 &
15.2$\pm$3.2 & 170 &  0.016 \nl
  & (+05.2, +00 10) & ~11.2$\pm$0.6 & 29$\pm$2.0 &
19.7$\pm$1.5 & 593 &  0.049 \nl
G10.30$-$0.15 & (+04.2, +00 50) & ~15.3$\pm$1.2  & 
16$\pm$2.6 & 14.6$\pm$2.8 & 237  & 0.063  \\
G23.46$-$0.20 & ($-$00.8, $-$02 00) & 104.1$\pm$1.5 & 
14$\pm$2.2 & 18.7$\pm$3.5 & 262  & 0.080  \\
G29.96$-$0.02 & (~~00.0, ~~~$\!$00 00) & ~97.1$\pm$0.7 &
13$\pm$0.8 & 21.4$\pm$1.7 & 254 &  0.068\\
  & ($-$00.6, $-$03 15) & 103.5$\pm$0.8 & 12$\pm$1.0 &
21.3$\pm$1.9 & 283 &  0.076 \\
  & (+05.6, $-$01 55) & 104.9$\pm$0.9 & 16$\pm$1.9 &
16.6$\pm$2.2 & 250 & 0.069 \\
\enddata
\tablenotetext{a}{Same as in Table 5}
\tablenotetext{}{NOTE. ~Among the 18 observed positions (Table 5),
He76$\alpha$ line emission was not detected (\tl$\lesssim$6~mK)
at 8 positions}
\end{deluxetable}

\clearpage

\begin{deluxetable}{ccccccccl}
\footnotesize
\tablewidth{0pt}
\tablecaption{RADIO RECOMBINATION AND MOLECULAR LINES OF \uchii\ REGIONS}
\tablehead{
  & \multicolumn{2}{c}{Compact Component\tablenotemark{a}} & & 
\multicolumn{4}{c}{\uchii\ Region}   &
\\ \cline{2-3} \cline{5-8}
  & \colhead{\vlsr} & \colhead{\delv} & & 
\colhead{RRL} &
\colhead{CS (7$-$6)} & \colhead{NH$_3$ (2,2)} & \colhead{NH$_3$ (4,4)} & 
\\
\colhead{Source} & \colhead{(\kms)} & \colhead{(\kms)} & &
\colhead{(\kms)} & 
\colhead{(\kms)} & \colhead{(\kms)} & \colhead{(\kms)} & Reference 
}
\startdata
~G5.89$-$0.39 & ~~9.9 & 54.8 & & ~~4.9 & ~~9.8 & ~~9.3 & ~11.0 & 
1, 5, 6, 7\\
~G5.97$-$1.17 & ~~4.5 & 26.8 & &\nodata & \nodata & ~~9.6 &\nodata& 
6\\
~G6.55$-$0.10 & ~13.2 & 26.5 &&$\sim$15& \nodata & ~12.8&\nodata & 
2, 6\\
~G8.14+0.23   & ~20.3 & 27.3 && \nodata & ~18.6 & ~19.3 & ~20.3 &
5, 6\\
G10.15$-$0.34 & ~16.4 & 29.7 && \nodata & \nodata & ~~9.9 & \nodata &
6\\
G10.30$-$0.15 & ~~7.7 & 28.5 && \nodata & \nodata & ~13.2 & ~12.7&
6, 7 \\
G12.21$-$0.10 & ~28.9 & 22.2 && ~27.6 & ~23.7 & ~24.3 & ~24.7 & 
3, 5, 6, 7\\
G12.43$-$0.05 & \nodata & \nodata && \nodata &\nodata & ~20.8 &
\nodata & 6\\
G23.71+0.17   & 103.0 & 30.3 && \nodata & \nodata & 113.0 & 114.6 &
6, 7\\
G23.96+0.15   & ~78.9 & 28.0 && ~73.7 & ~80.4 & ~79.7 & \nodata &
1, 5, 6\\
G27.28+0.15   & ~35.0 & 23.9 &&\nodata &\nodata & ~31.5 & \nodata &
6\\
G29.96$-$0.02 & ~95.9 & 28.6 && ~95.6 & ~97.7 & ~97.6 & 98.1 &
4, 5, 6, 7\\
G35.05$-$0.52 & ~51.2 & 17.9 && \nodata & \nodata & ~49.4 & \nodata &
6\\
G37.55$-$0.11 & ~51.5 & 30.3 &&\nodata &\nodata &
$\sim$52\tablenotemark{b} &\nodata & 6\\
\hline
$< v_{\rm LSR,UC}-v_{\rm LSR,C} >$ & & & & $-$2.0$\pm$3.0 &
$-$0.6$\pm$2.9& 0.4$\pm$4.5 & 2.7$\pm$5.4 & \\
\enddata
\tablenotetext{a}{\hrrl\ data from this work}
\tablenotetext{b}{NH$_3$ (1,1) data from Churchwell et al. (1990a)}
\tablenotetext{}{REFERENCES.---(1) \hrrl\ data from Churchwell et al. 1990b. (2)
\hrrl\ data from Andrews et al. 1985. (3) H93$\alpha$ data from
Afflerbach et al. 1996. (4) \hrrl\ data from Afflerbach et al. 1994. (5)
CS (7$-$6) data from Plume et al. 1992. (6) NH$_3$ (2,2) data from
Churchwell et al. 1990a. (7) NH$_3$ (4,4) data from Cesaroni et al. 1992.}
\end{deluxetable}

\clearpage
 

\begin{figure}[tbp]

\vskip -0.7cm

\begin{minipage}[h]{.46\linewidth}
\centering\epsfxsize=3.2in\epsfbox{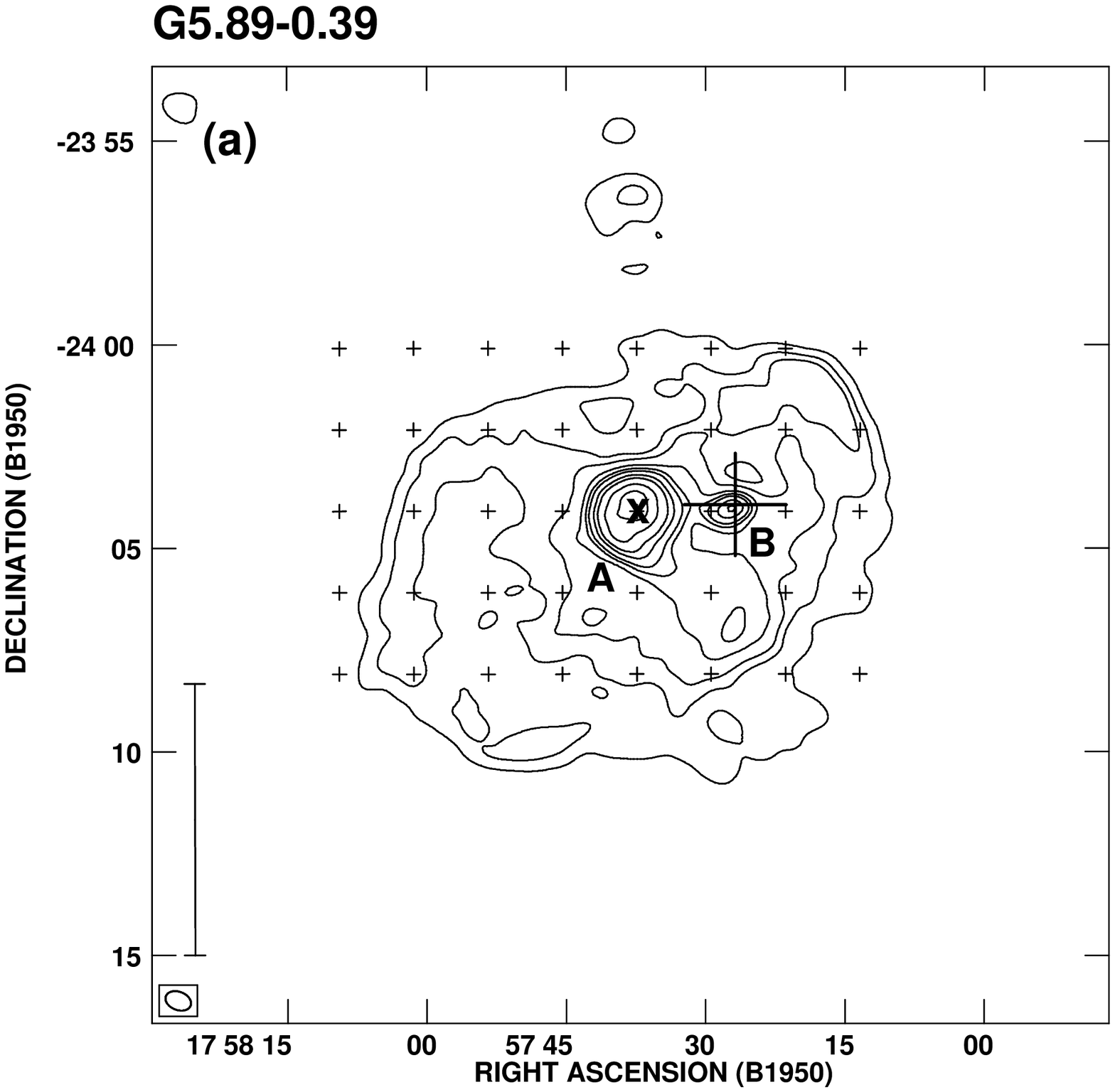}
\end{minipage}\hfill
\begin{minipage}[h]{.46\linewidth}
\centering\epsfxsize=3.2in\epsfbox{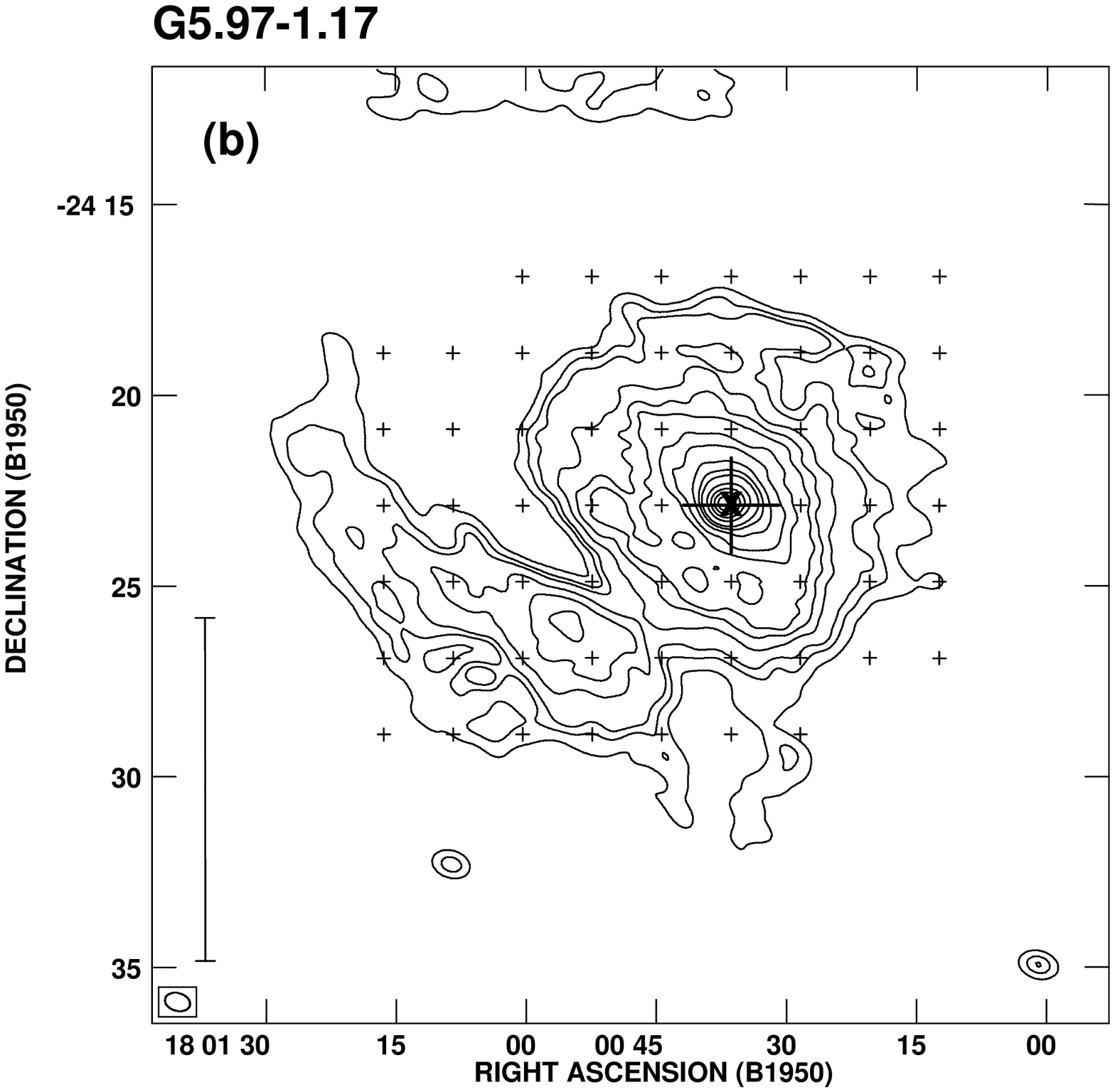}
\end{minipage}

\vskip -0.5cm
\begin{minipage}[h]{.46\linewidth}
\centering\epsfxsize=3.2in\epsfbox{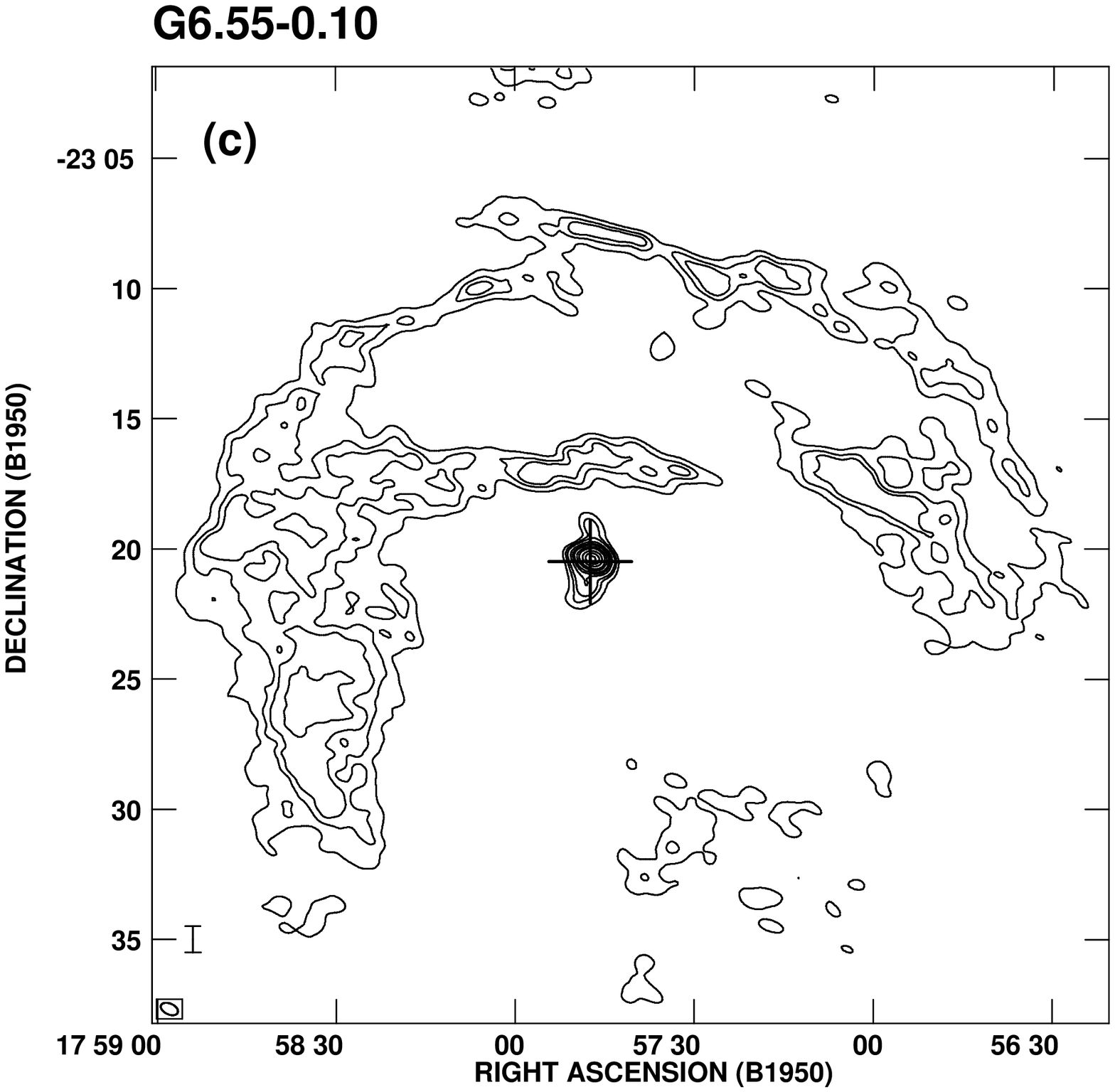}
\end{minipage}\hfill
\begin{minipage}[h]{.46\linewidth}
\centering\epsfxsize=3.2in\epsfbox{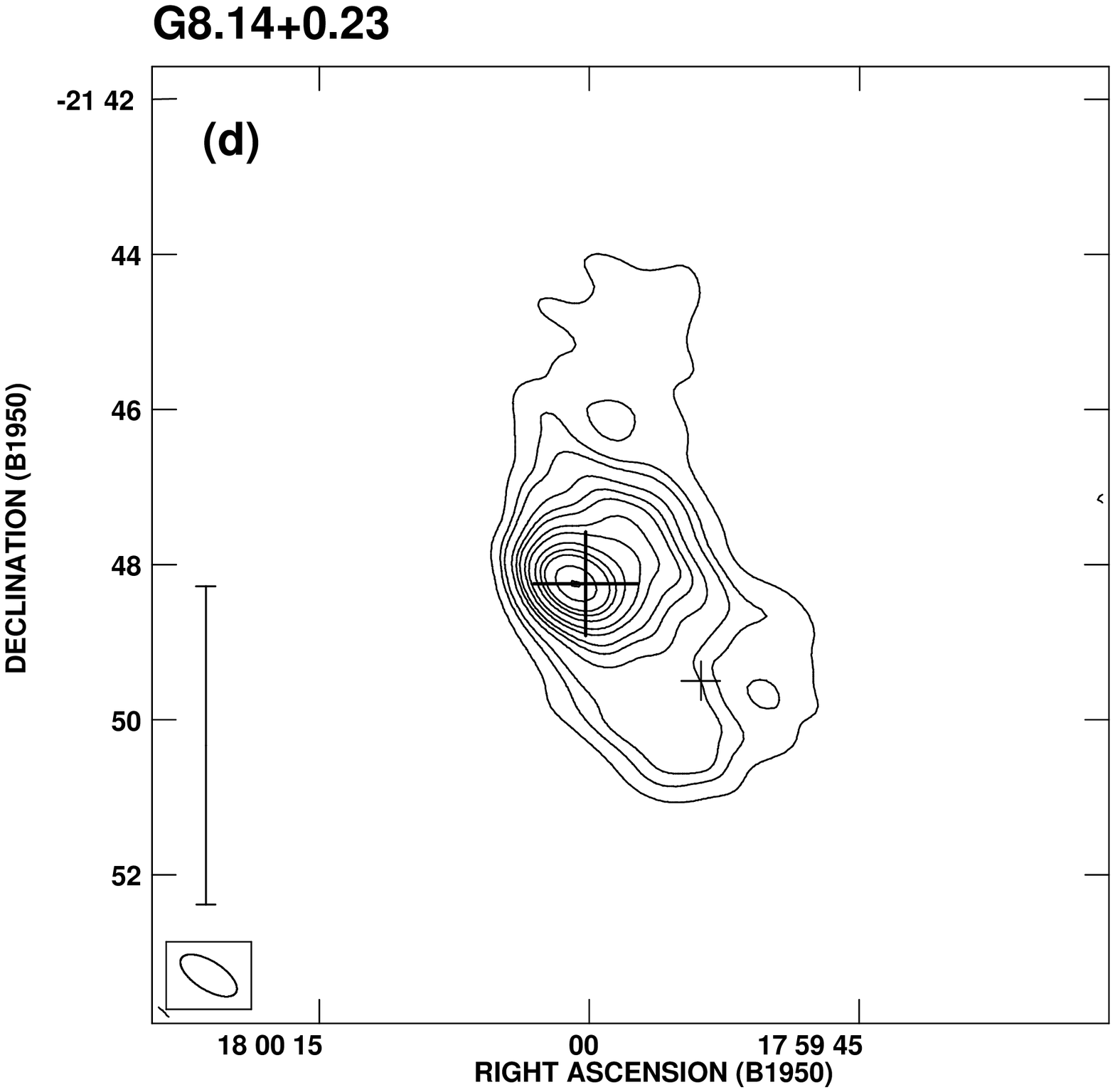}
\end{minipage}

\vskip -0.5cm
\begin{minipage}[h]{.46\linewidth}
\centering\epsfxsize=3.2in\epsfbox{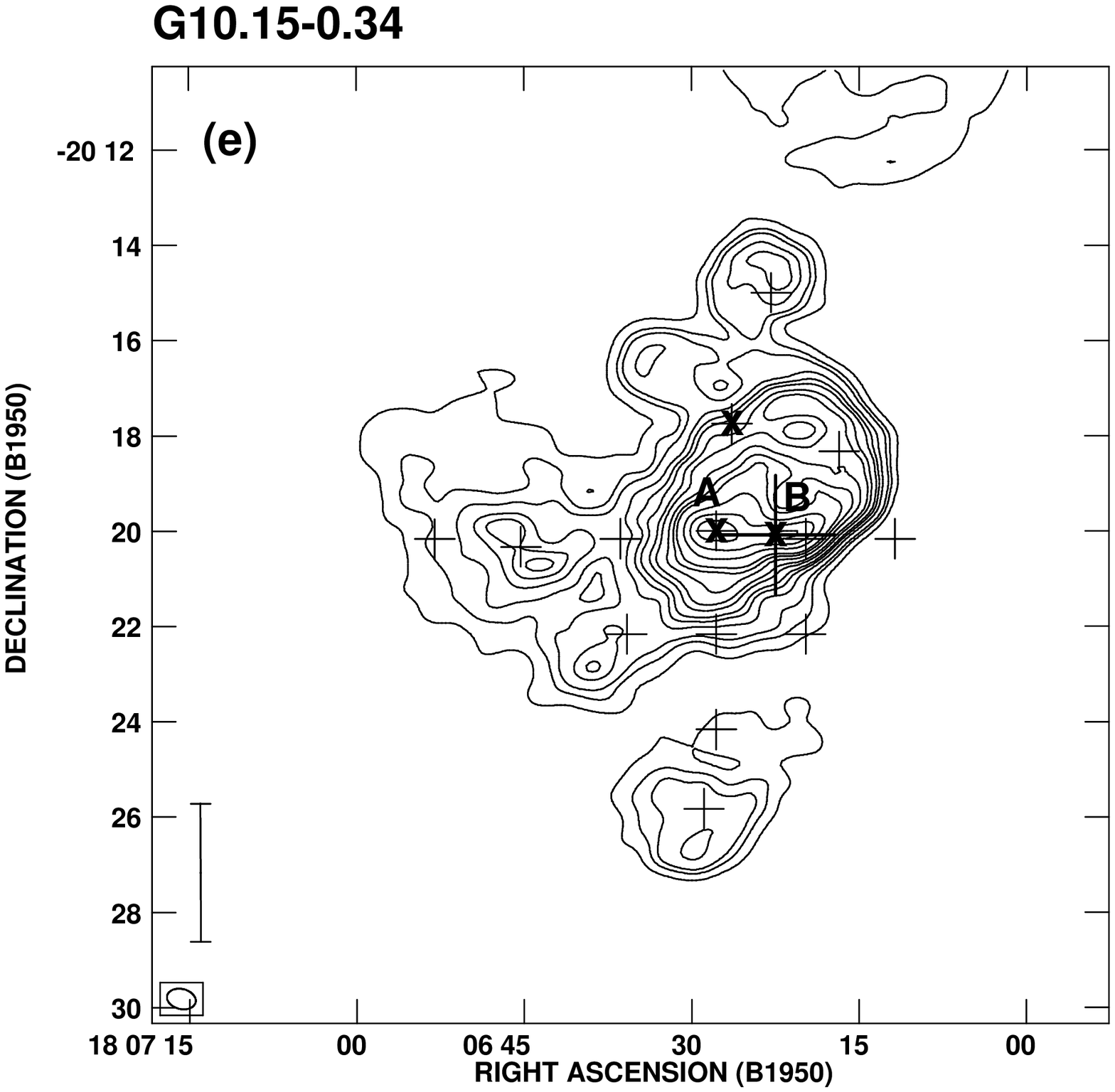}
\end{minipage}\hfill
\begin{minipage}[h]{.46\linewidth}
\centering\epsfxsize=3.2in\epsfbox{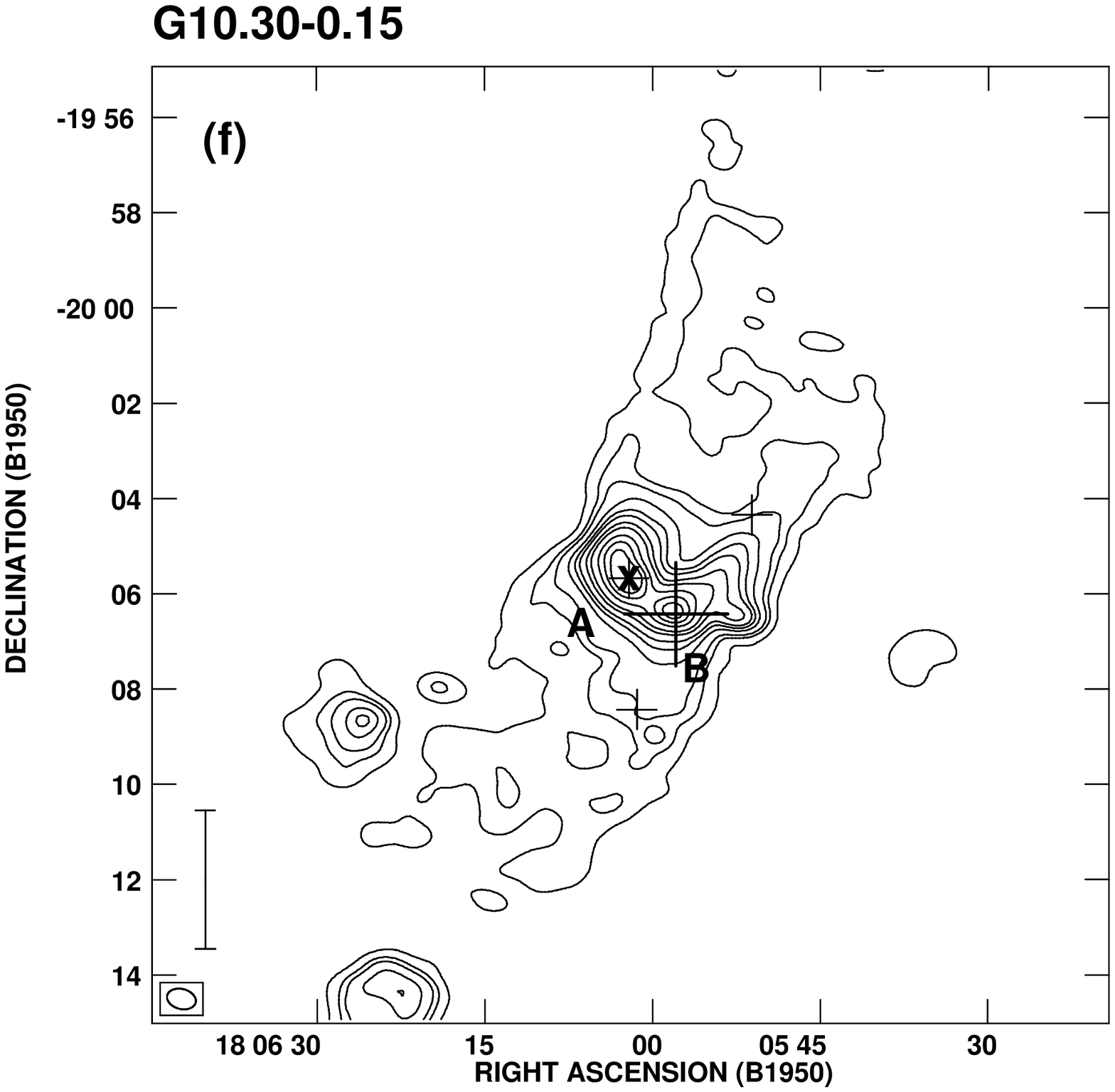}
\end{minipage}

\end{figure}
 

\begin{figure}[tbp]

\hskip -0.5cm
\vskip -0.7cm
\begin{minipage}[h]{.46\linewidth}
\centering\epsfxsize=3.2in\epsfbox{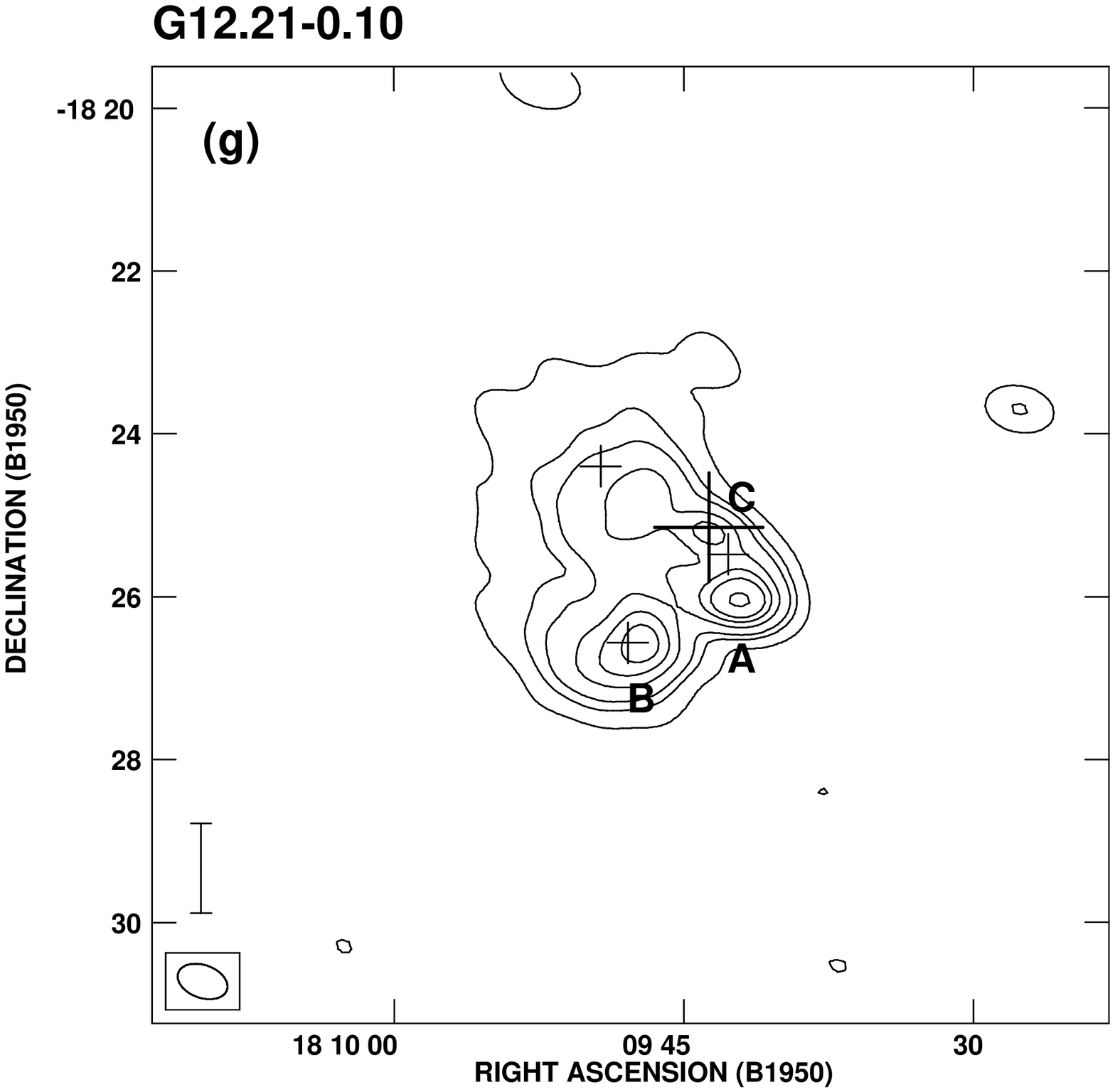}
\end{minipage}\hfill
\begin{minipage}[h]{.46\linewidth}
\centering\epsfxsize=3.2in\epsfbox{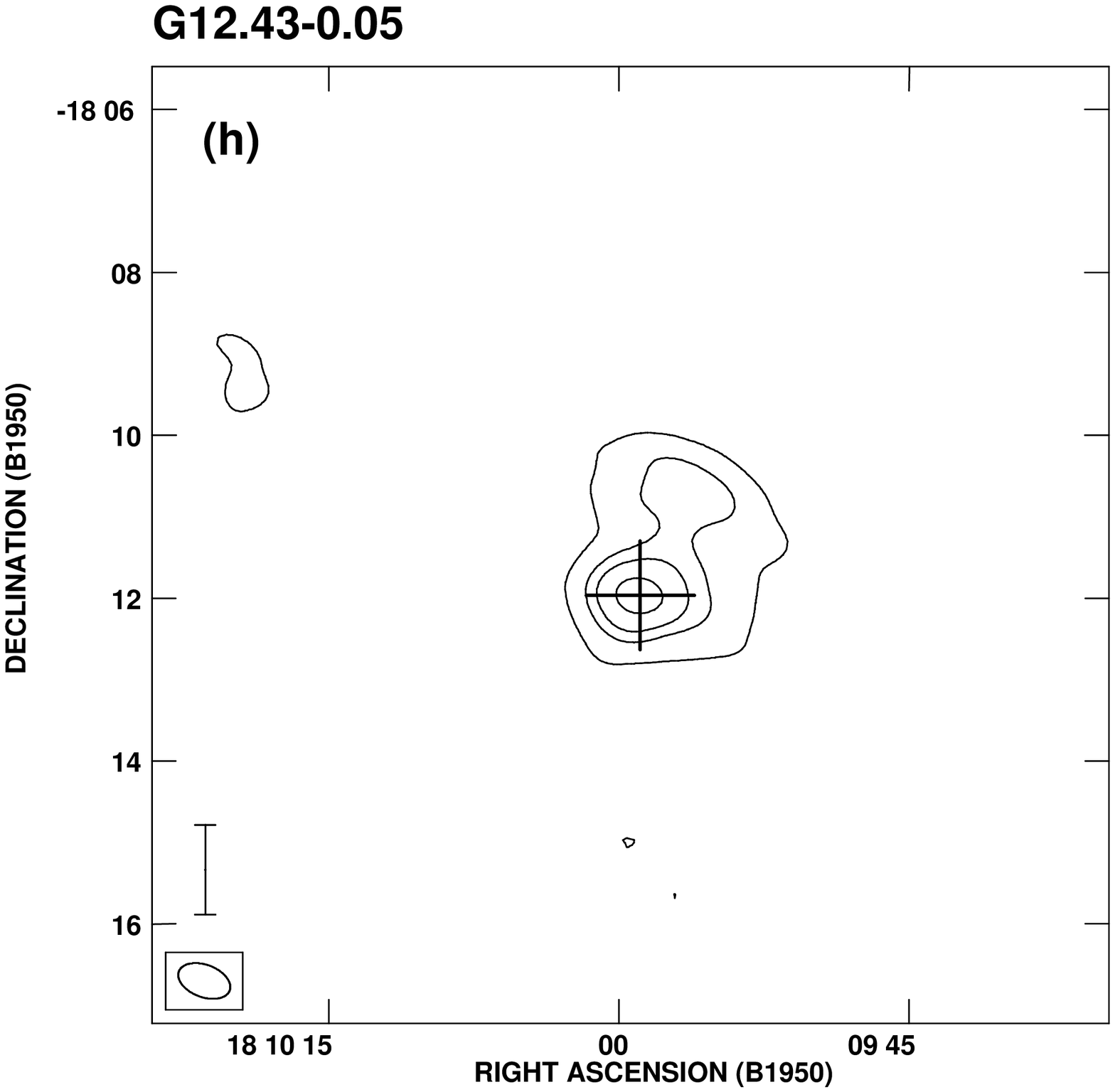}
\end{minipage}
 
\vskip -0.5cm
\begin{minipage}[h]{.46\linewidth}
\centering\epsfxsize=3.2in\epsfbox{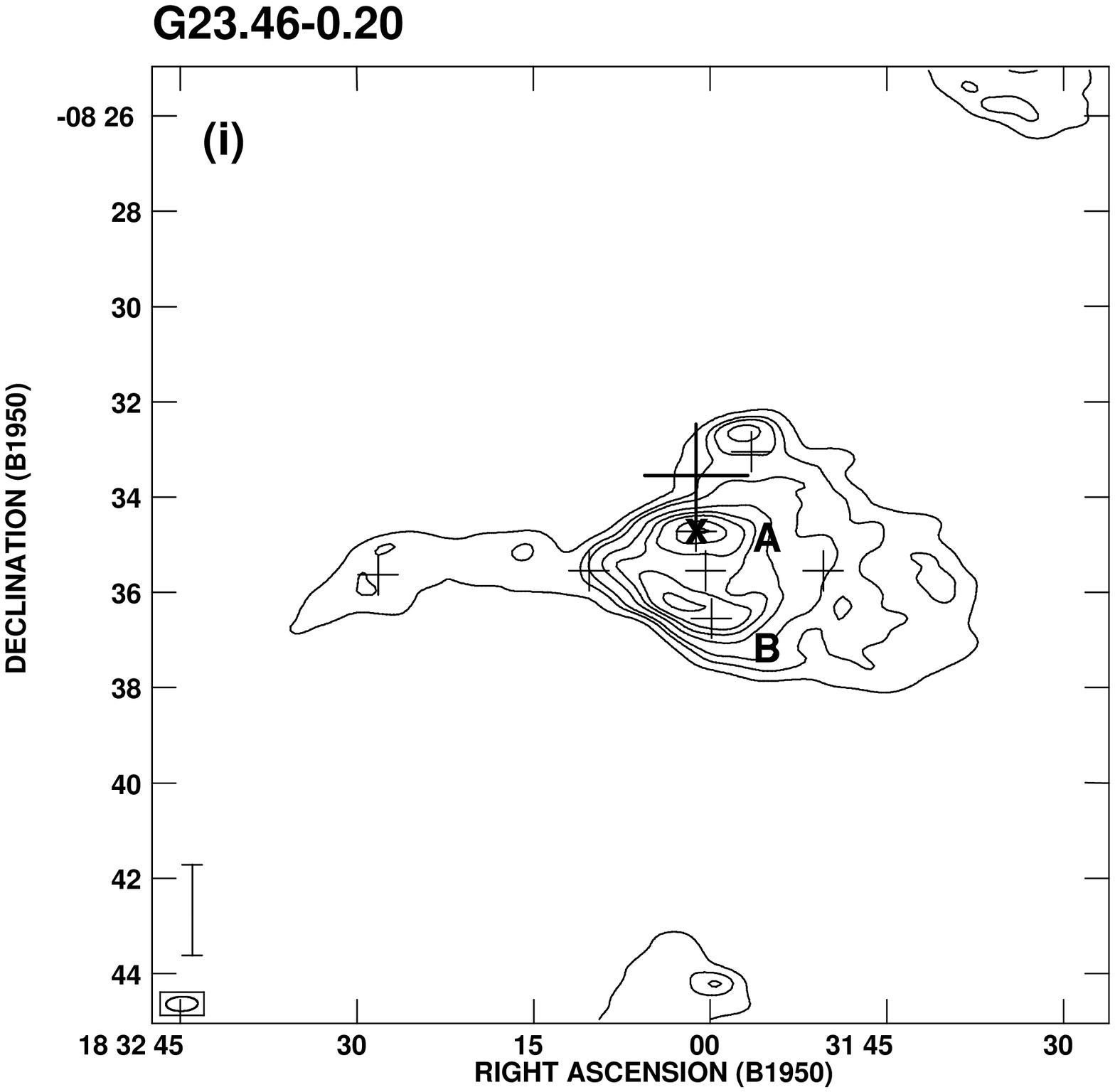}
\end{minipage}\hfill
\begin{minipage}[h]{.46\linewidth}
\centering\epsfxsize=3.2in\epsfbox{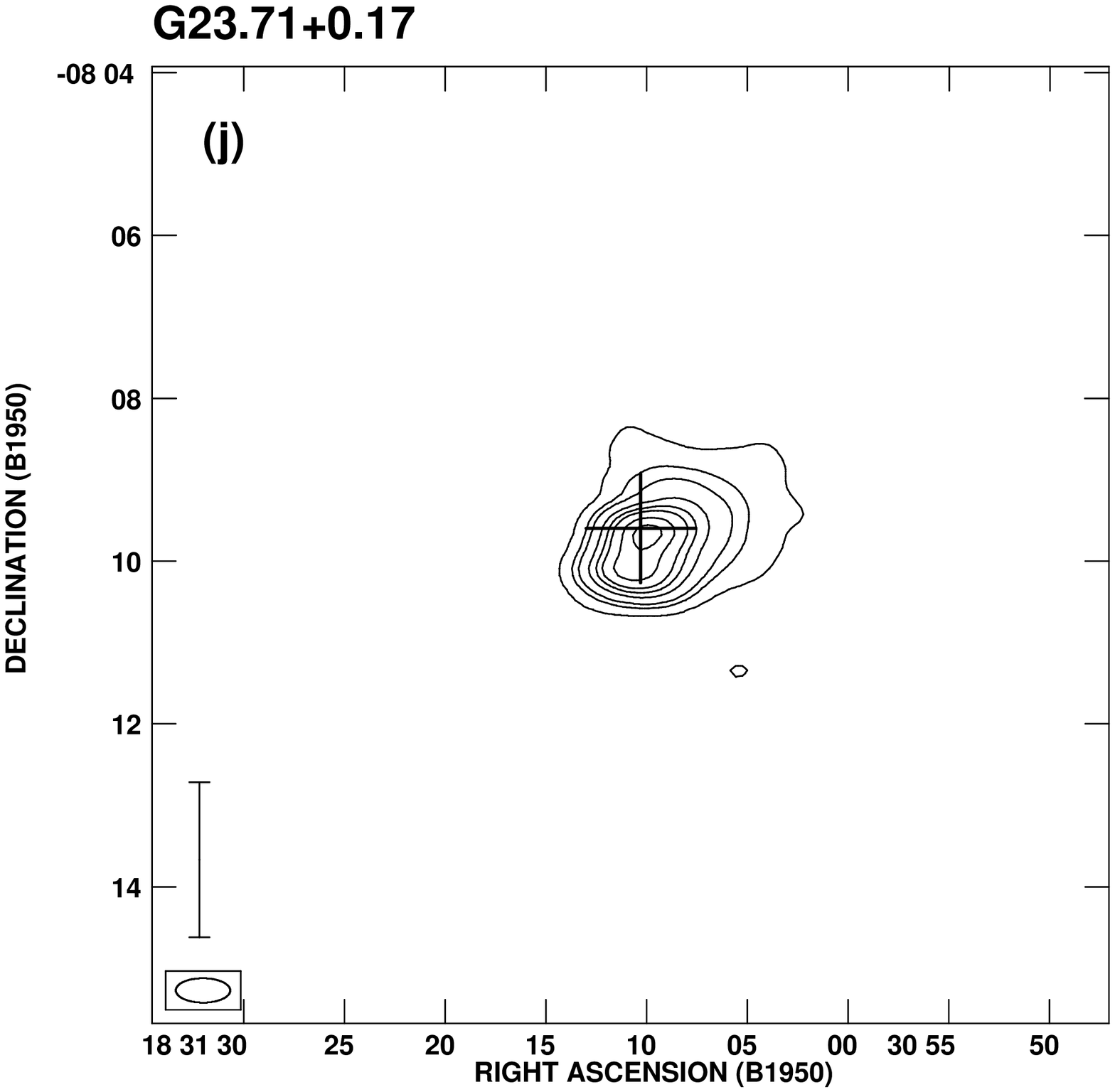}
\end{minipage}
 
\vskip -0.5cm
\begin{minipage}[h]{.46\linewidth}
\centering\epsfxsize=3.2in\epsfbox{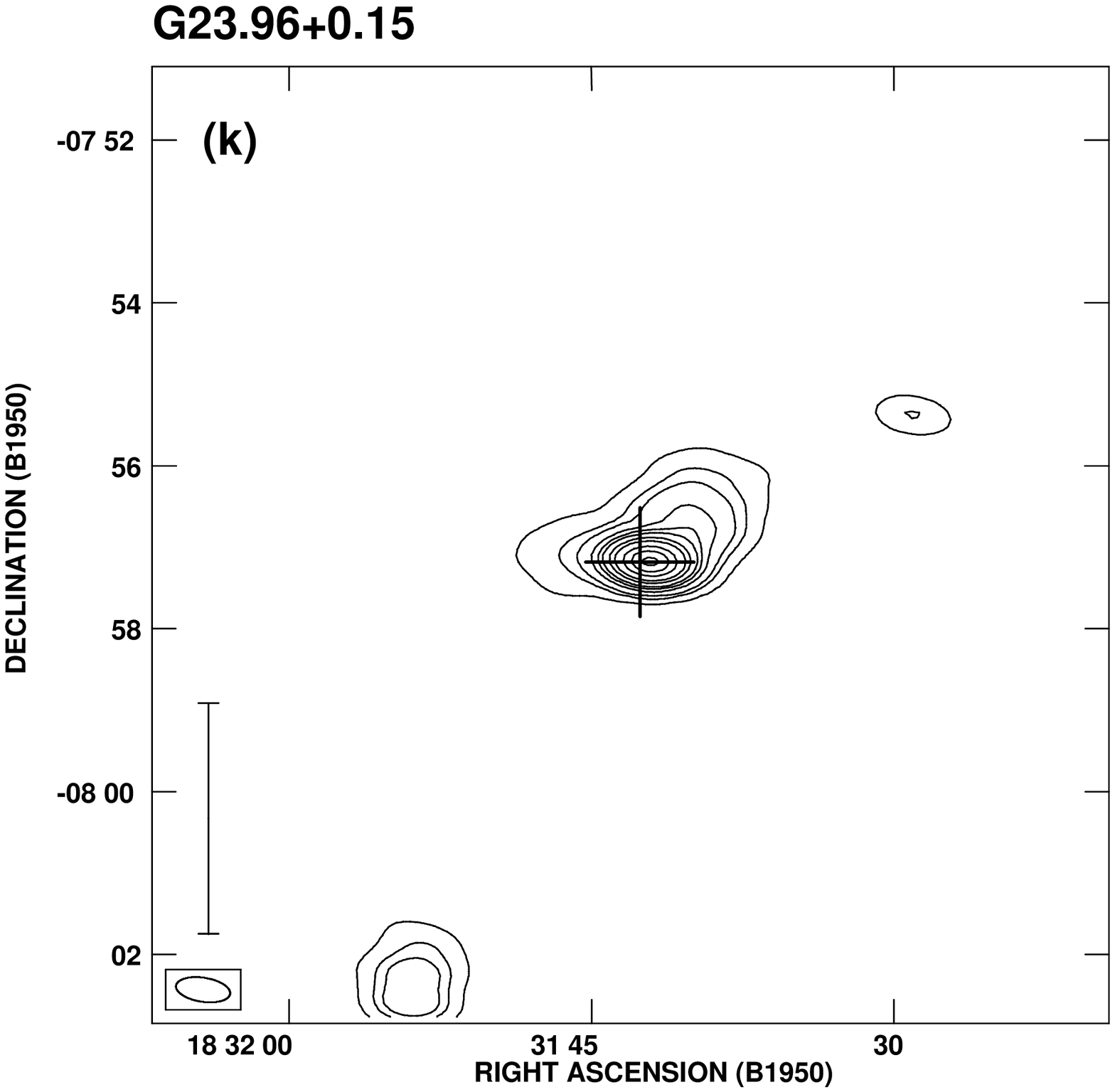}
\end{minipage}\hfill
\begin{minipage}[h]{.46\linewidth}
\centering\epsfxsize=3.2in\epsfbox{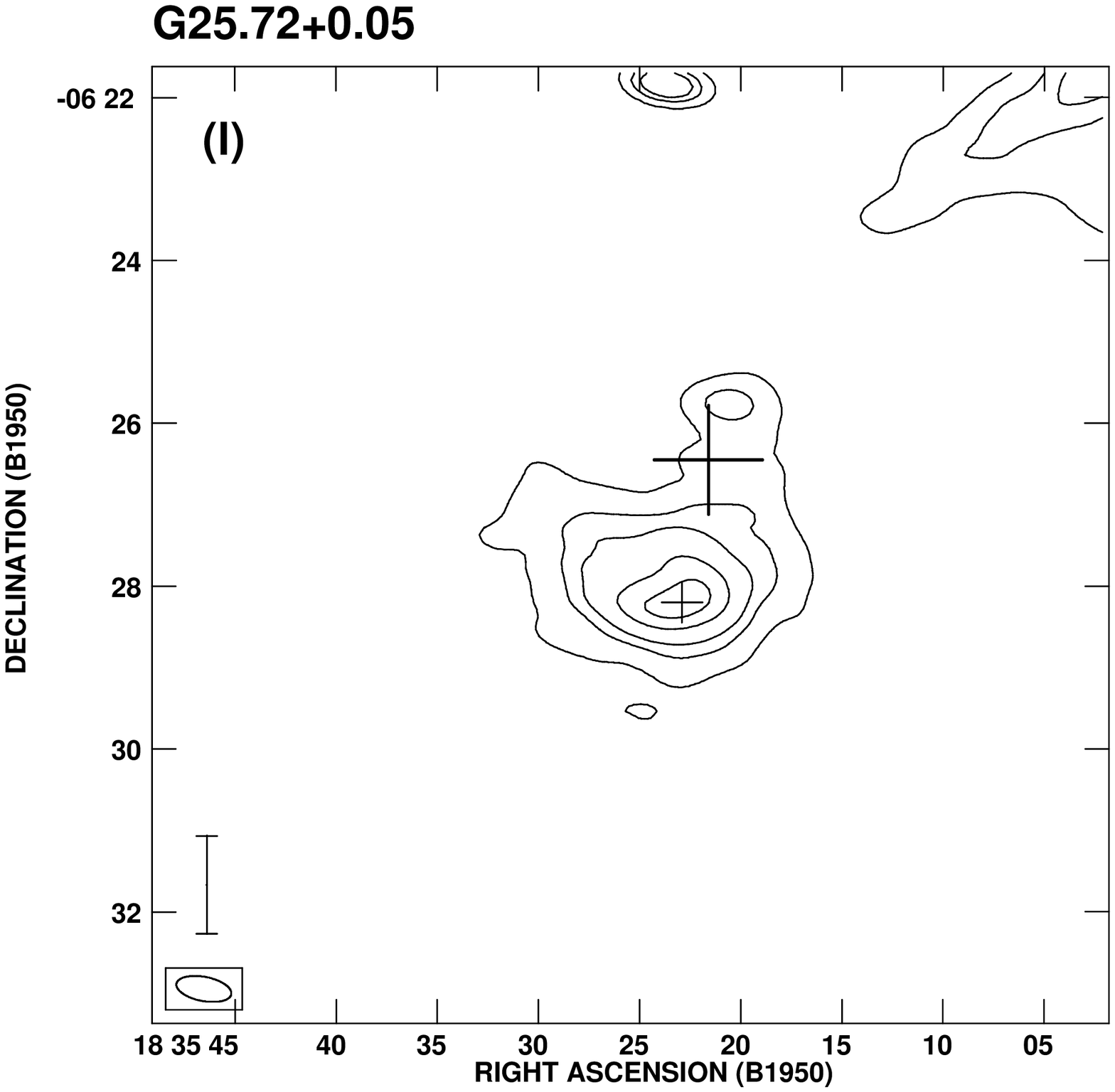}
\end{minipage}

\end{figure}


\begin{figure}[tbp]

\vskip -2.0cm
\begin{minipage}[h]{.46\linewidth}
\centering\epsfxsize=3.2in\epsfbox{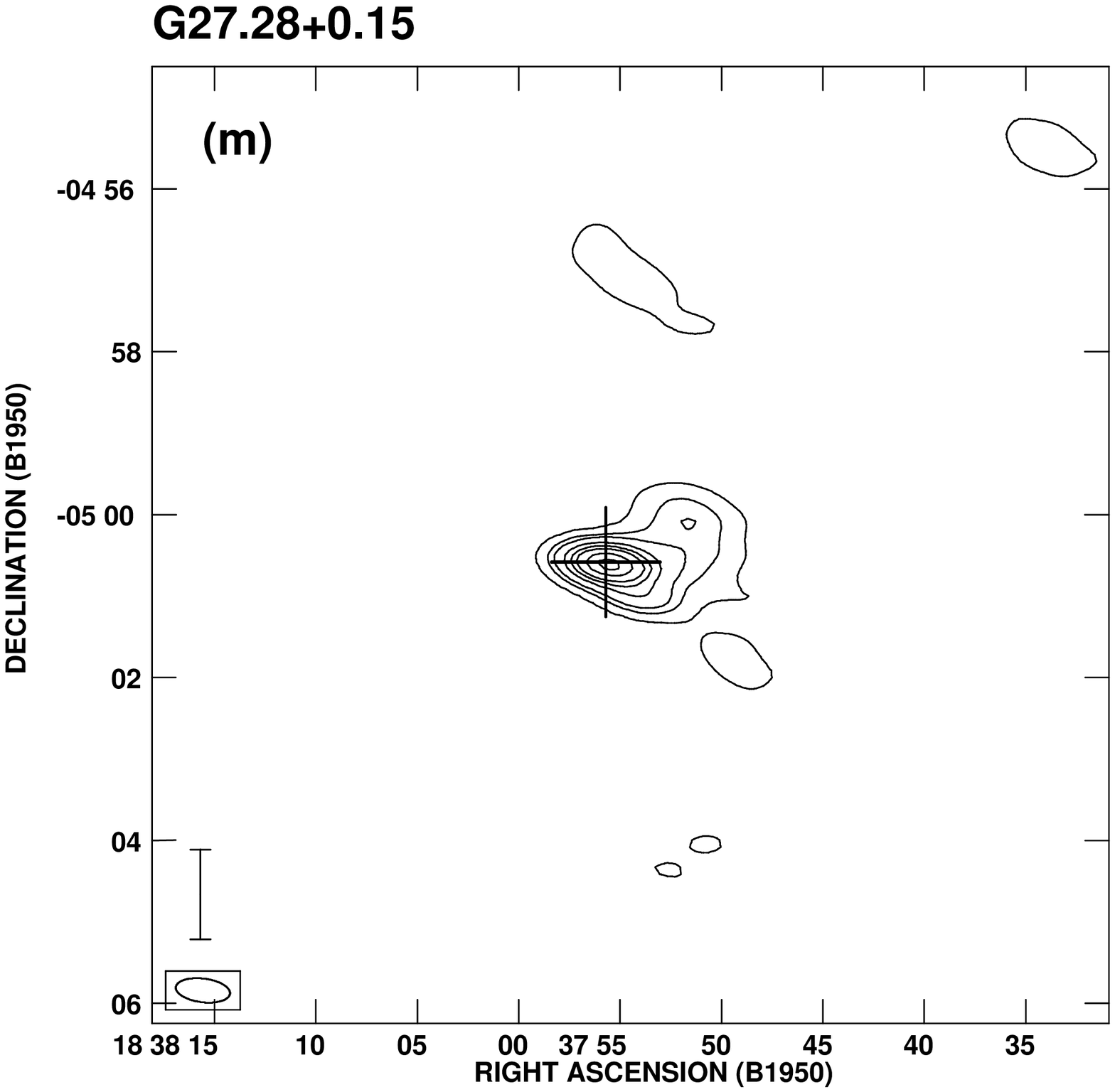}
\end{minipage}\hfill
\begin{minipage}[h]{.46\linewidth}
\centering\epsfxsize=3.2in\epsfbox{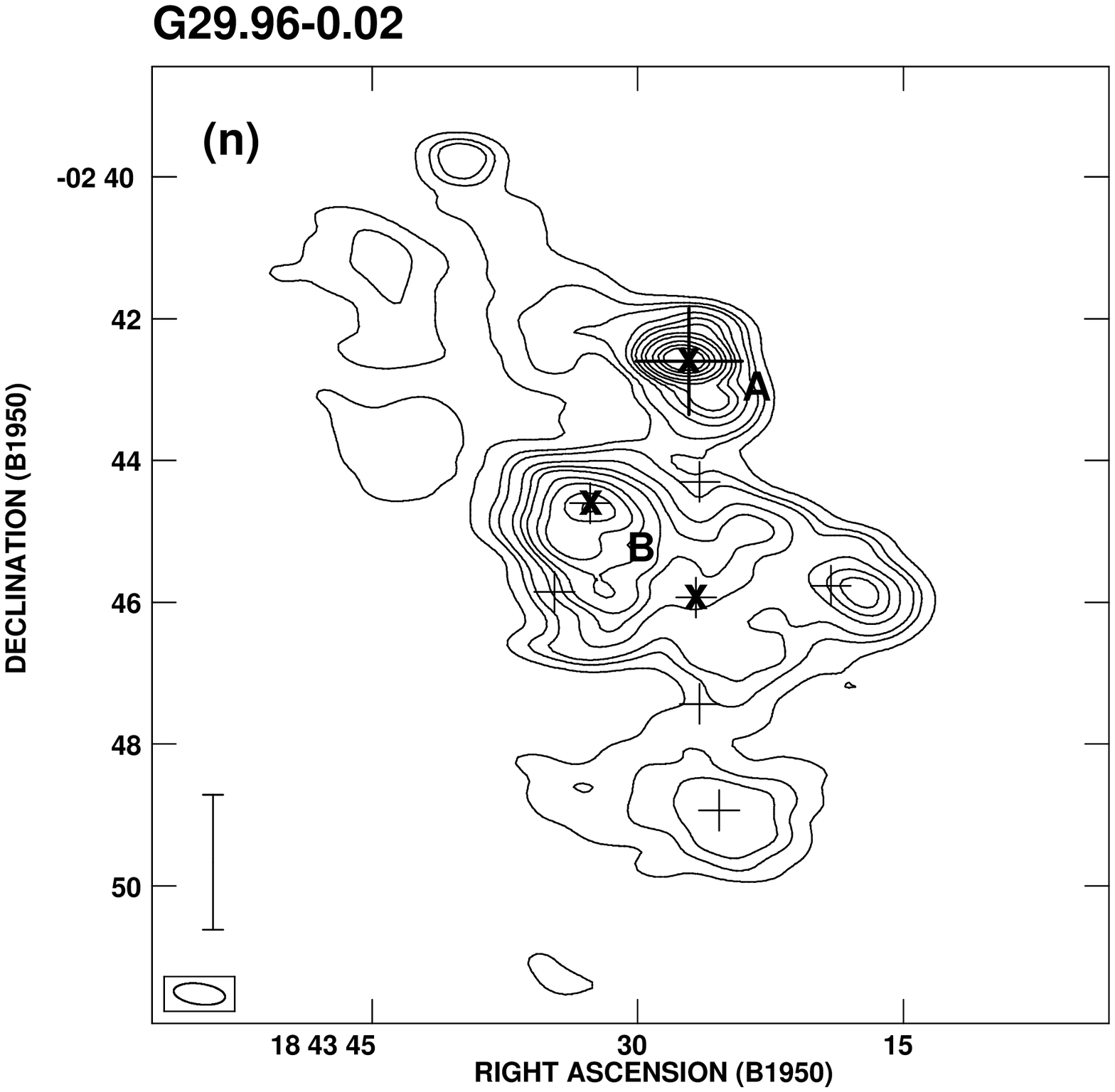}
\end{minipage}
 
\vskip -0.5cm
\begin{minipage}[h]{.46\linewidth}
\centering\epsfxsize=3.2in\epsfbox{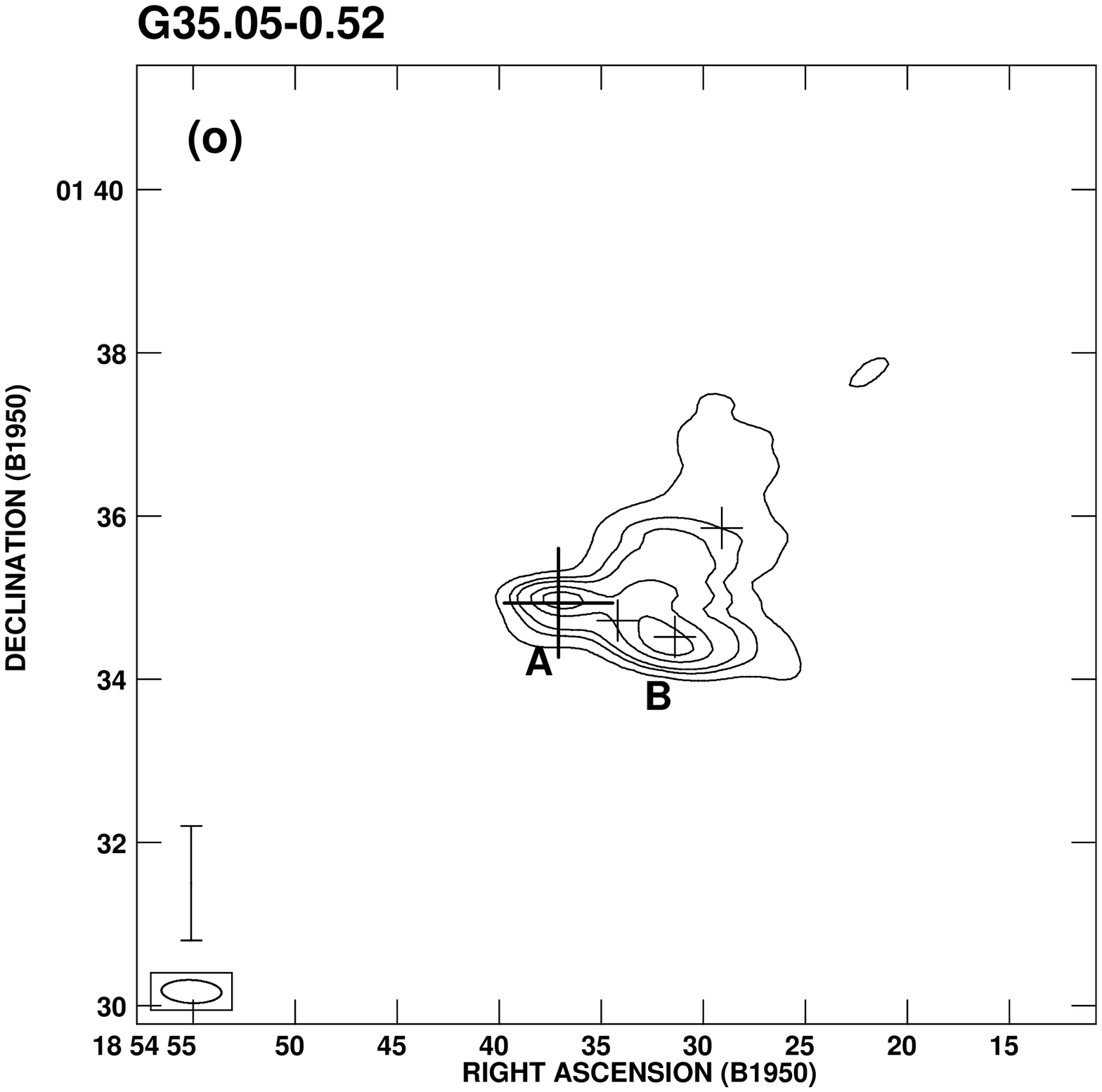}
\end{minipage}\hfill
\begin{minipage}[h]{.46\linewidth}
\centering\epsfxsize=3.2in\epsfbox{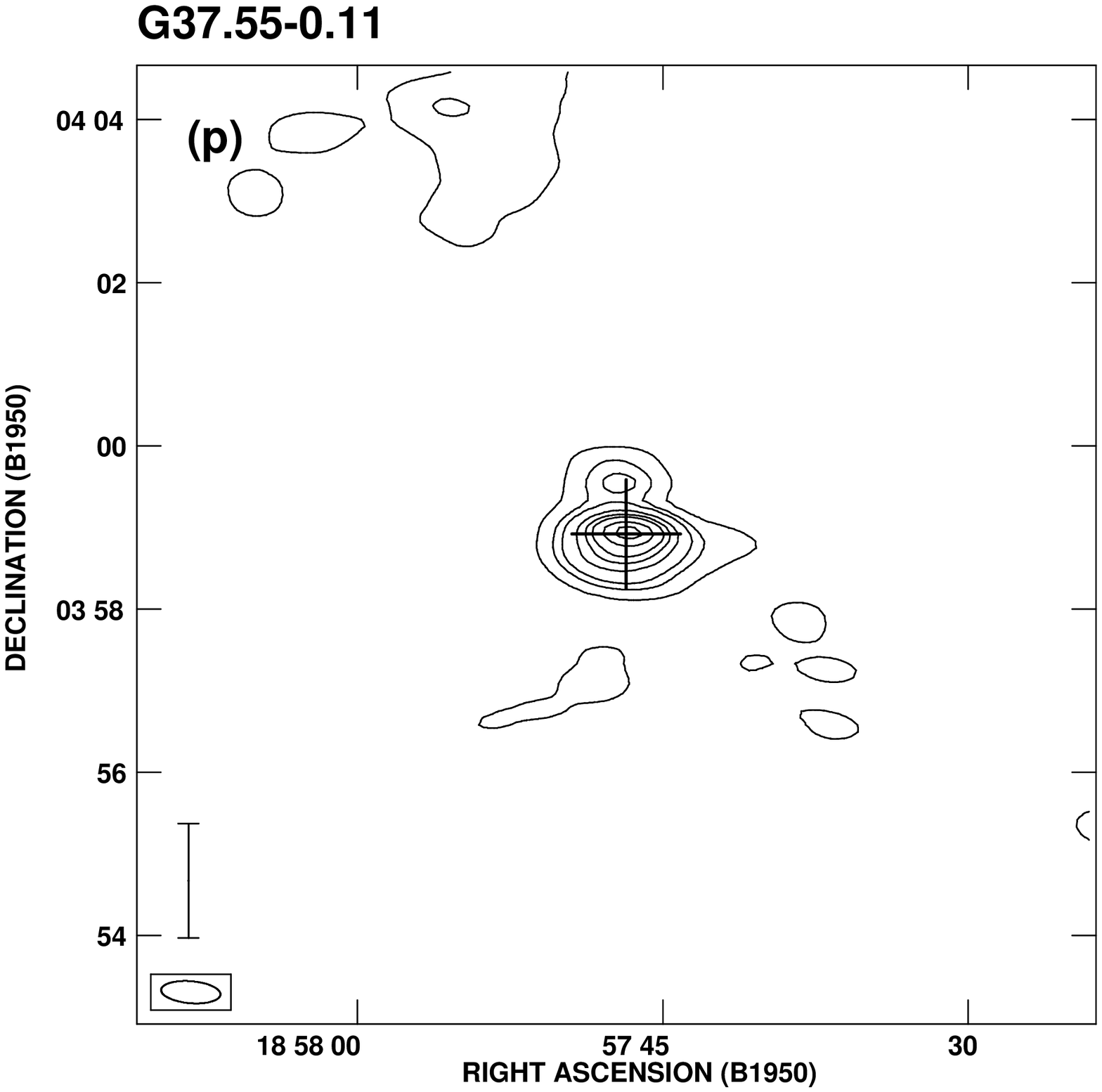}
\end{minipage}

\figcaption[figure1.ps]{
Radio continuum maps of 16 \uchii\ regions,
made with the VLA (DnC array) at 21~cm.
Contour levels are 10, 30, 50, 100, 150, 200, 300, 400, 600,
800, 1000, 1500, 2000, 2500, and 3000~mJy~beam$^{-1}$.
In each field, a large cross
represents an \uchii\ region while small crosses are the positions
where the \hrrl\ line was observed.
The ``$\times$'' symbols mark the positions with detectable \herrl\ line
emission (Table 6). In the sources with two or more compact components,
A$-$C refer to the entries in Table 3.
The synthesized beam and a 5~pc linear scale bar are shown in the bottom
left corner. The linear scale assumes the distance given in Table 2.
}

\end{figure}

\clearpage

\begin{figure}
\vskip -5cm
\figurenum{2}
\epsscale{1.0}
\plotone{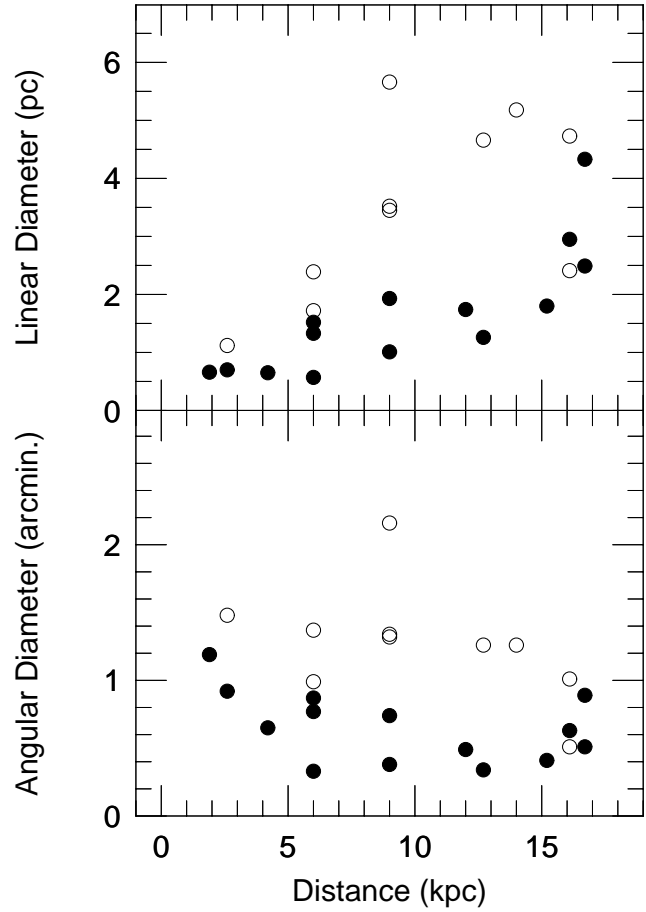}
\vskip -2cm
\figcaption[figure2.ps]{
Plot of angular ({\it lower panel}) and linear ({\it upper panel}) sizes
versus distance for the compact components.
Filled and open circles represent the compact components with and
without \uchii\ regions, respectively. The compact components with
\uchii\ regions are generally smaller than those without \uchii\ regions.
}
\end{figure}
 
\clearpage

\begin{figure}
\vskip -6cm
\figurenum{3}
\epsscale{1.0}
\plotone{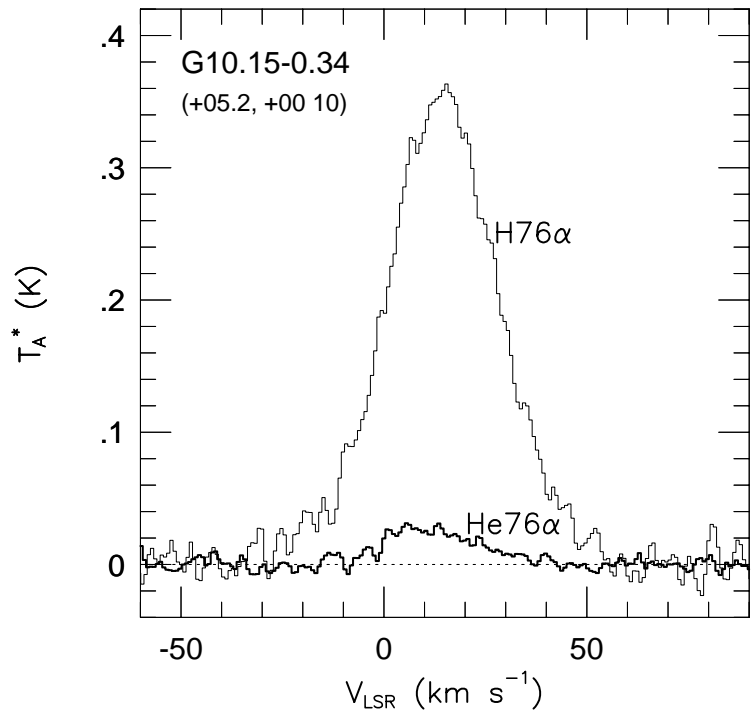}
\vskip -3cm
\figcaption[figure3.ps]{
\hrrl\ and \herrl\ radio recombination lines at G10.15$-$0.34
(+05.$^{\!\!\rm s}$2, +00$'$ 10$''$), i.e., $(\alpha, \delta)_{1950}$=
(18$^{\rm h}$ 06$^{\rm m}$ 27.$^{\!\!\rm s}$7, $-20^\circ$ 19$'$ 55$''$),
where the strongest \hrrl\ line emission was detected
(Tables 5 \& 6).
}
\end{figure}
 
\clearpage

\begin{figure}
\vskip -5cm
\figurenum{4}
\epsscale{1.0}
\plotone{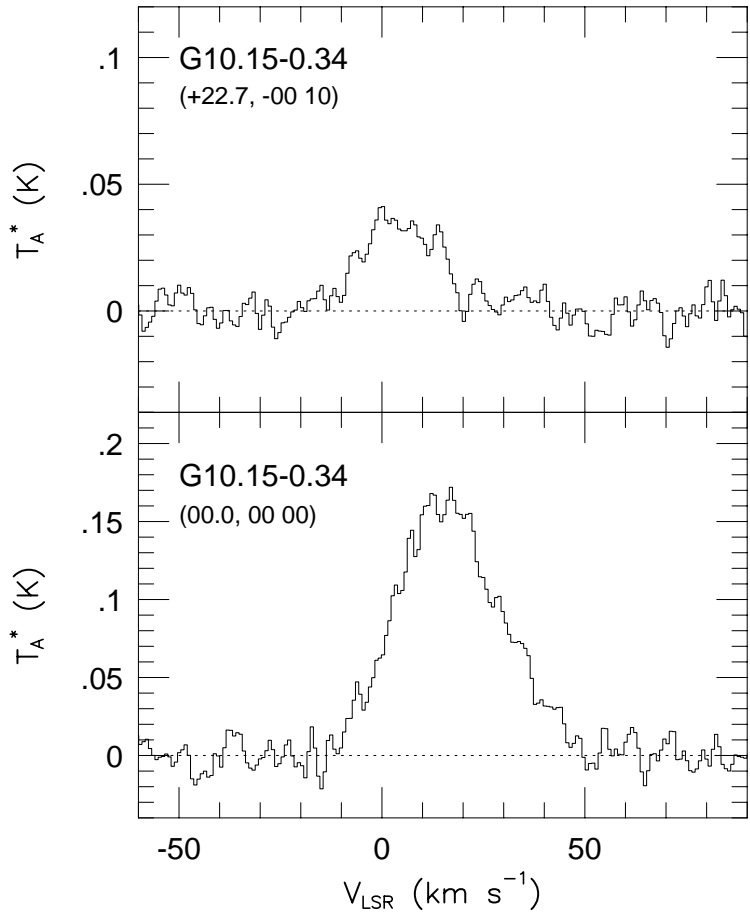}
\vskip -3cm
\figcaption[figure4.ps]{
\hrrl\ line profiles at G10.15$-$0.34
(+22.$^{\!\!\rm s}$7, $-$00$'$ 10$''$) and (00.$^{\!\!\rm s}$0, 00$'$ 00$''$).
The former is blueshifted by about 10~\kms\ with respect to
the latter (Table 5). This suggests that the eastern protuberance
may be a result of the champagne flow.
}
\end{figure}
 
\clearpage

\begin{figure}
\vskip -6cm
\figurenum{5}
\epsscale{1.0}
\plotone{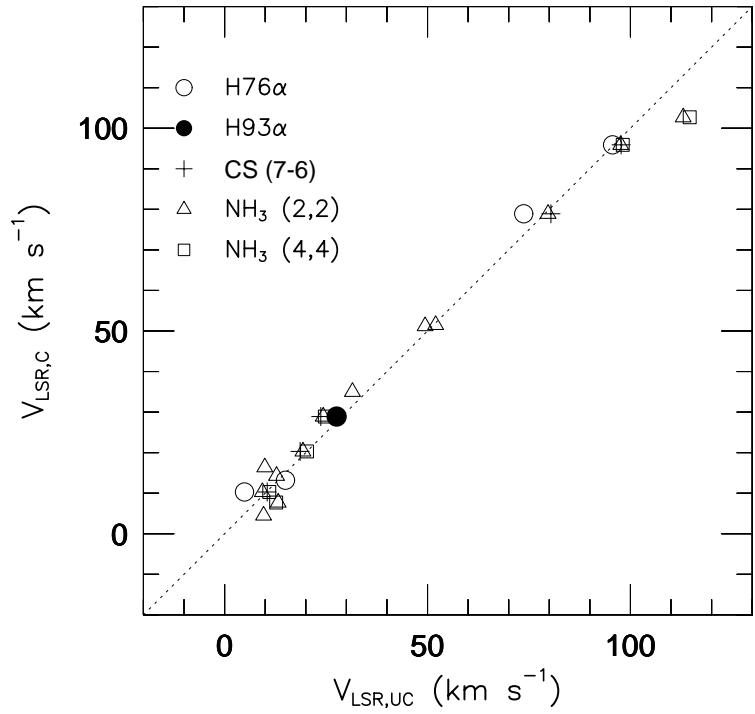}
\vskip -2cm
\figcaption[figure5.ps]{
The central velocities of \uchii\ regions are compared with those of
their associated compact components.
\hrrl, H93$\alpha$, CS (7$-$6), NH$_3$ (2,2), and NH$_3$ (4,4) lines
are marked by
open circles, filled circles, crosses, triangles, and squares, respectively.
The dotted line represents $v_{\rm LSR,C}$=$v_{\rm LSR,UC}$.
}
\end{figure}
 
\clearpage

\begin{figure}
\vskip -9cm
\hskip -2cm
\figurenum{6}
\epsscale{1.1}
\plotone{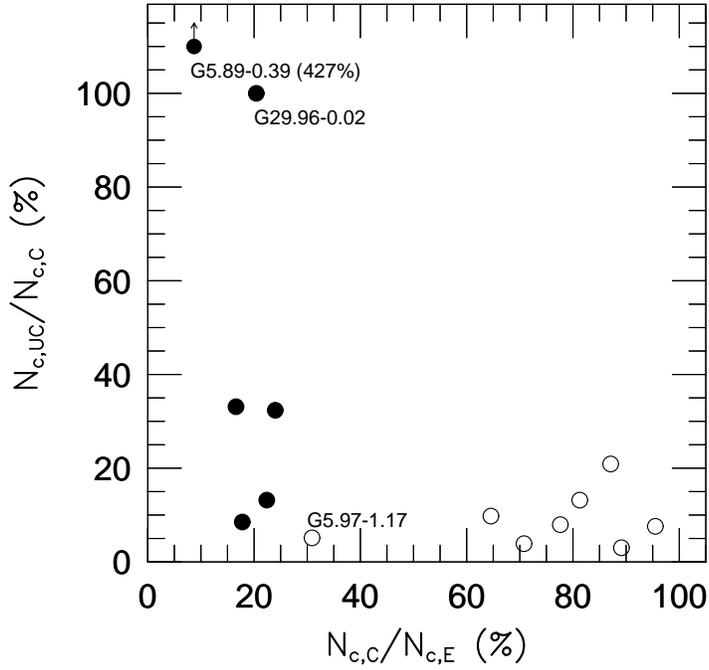}
\vskip -2.5cm
\figcaption[figure6.ps]{
Comparison of \ncuc/\ncc\ with \ncc/\nce\ for 14 sources in which
\uchii\ regions have associated compact components.
Here \ncuc, \ncc, and \nce\ are the Lyman continuum photon fluxes of
the ultracompact, compact, and extended components, respectively.
Open circles are sources with a single peak, while filled circles are
sources with two or more compact components.
For G5.89$-$0.39 and G29.96$-$0.02,
\ncc\ is equal to or less than \ncuc, probably due to a larger optical
depth at 21~cm.
}
\end{figure}
 
\clearpage

\begin{figure}
\vskip -8cm
\figurenum{7}
\epsscale{1.1}
\plotone{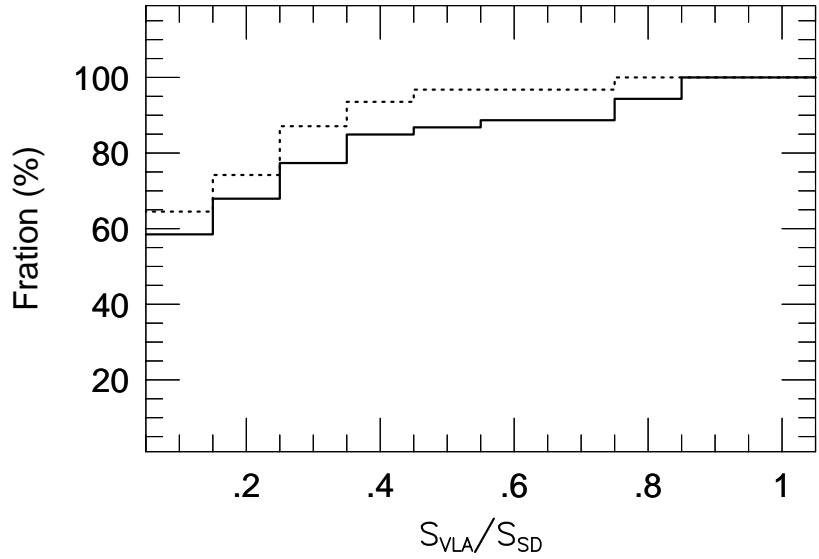}
\vskip -4cm
\figcaption[figure7.ps]{
Variation of fraction (\%) with
ratio of VLA to single-dish fluxes, S$_{\rm VLA}$/S$_{\rm SD}$.
The value at the ordinate is the number fraction of \uchii\ regions
with smaller flux ratios than a given value at the abscissa.
The solid line is drawn for 30 \uchii\ regions of simple morphology in the
catalog of WC89, while the dotted line is for 22 simple \uchii\ regions
in the catalog of KCW.
}
\end{figure}
 
\clearpage

\begin{figure}
\vskip -1cm
\figurenum{8}
\epsscale{1.0}
\plotone{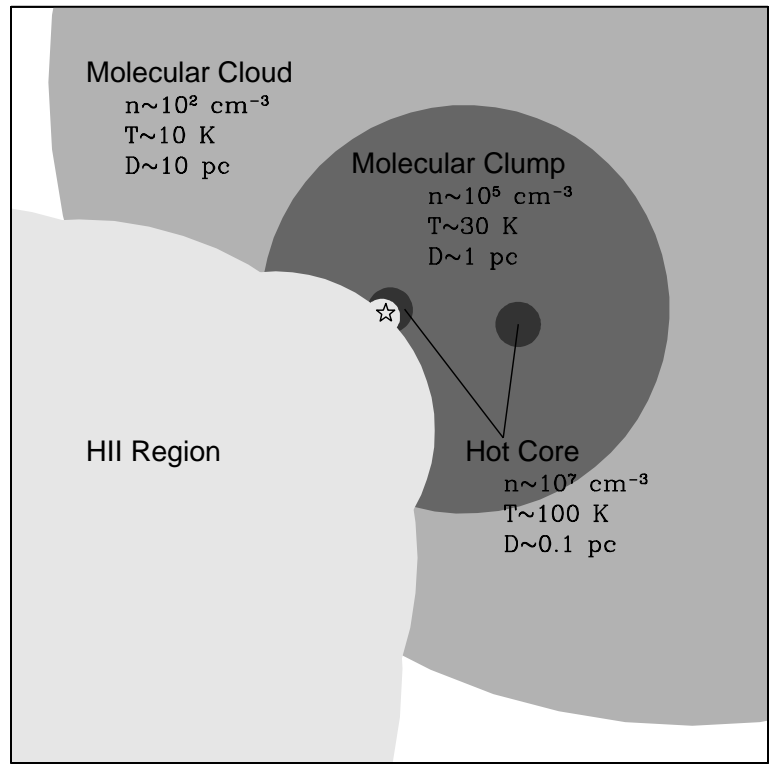}
\vskip -2cm
\figcaption[figure8.ps]{
Schematic representation of our model that explains the origin of
extended emission around an \uchii\ region using the Champagne flow
model and the hierarchical structure of molecular clouds.
An O star, marked by a star symbol, is
located off-center within the hot core embedded in a molecular clump.
The star maintains the ionization of the ultracompact, compact, and
extended \hii\ regions. This figure is not scaled.
See the text for more details.
}
\end{figure}
 
\clearpage

\begin{figure}
\vskip -5cm
\figurenum{9}
\epsscale{1.0}
\plotone{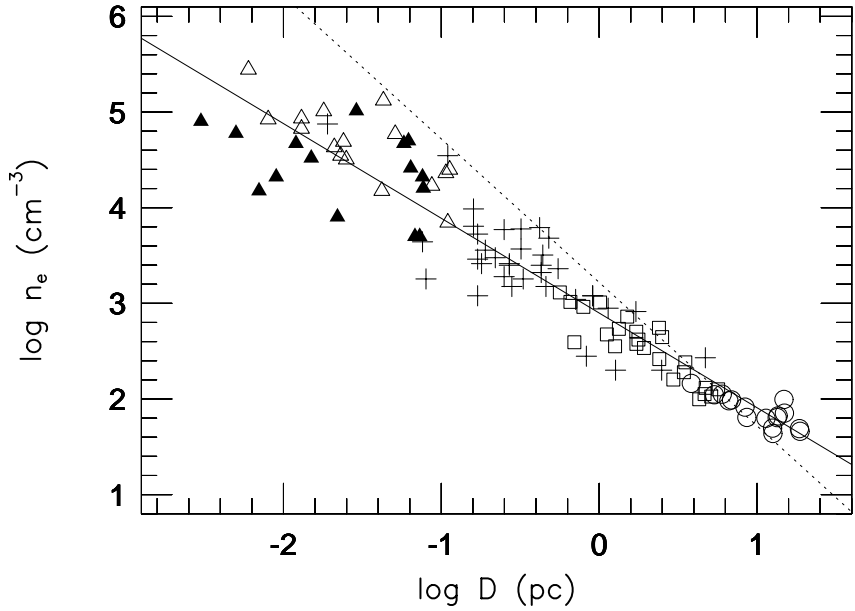}
\vskip -3cm
\figcaption[figure9.ps]{
Plot of \nee\ against D. Various \hii\ regions are marked:
open triangles ({\it spherical UC ones in WC89}),
filled triangles ({\it spherical UC ones in KCW}),
crosses ({\it UC and compact ones in Garay et al. (1993)}),
open squares ({\it compact ones in this work}),
and open circles ({\it extended ones in this work}).
A linear fit to all the data points gives \nee=790~D$^{-0.99}$ with a
correlation coefficient of $-$0.95. The solid and dotted lines represent
the fitted relation and the line of a constant excitation parameter ($U$=70),
respectively.
}
\end{figure}
 
\clearpage

\begin{figure}
\vskip -1cm
\figurenum{10}
\epsscale{0.7}
\plotone{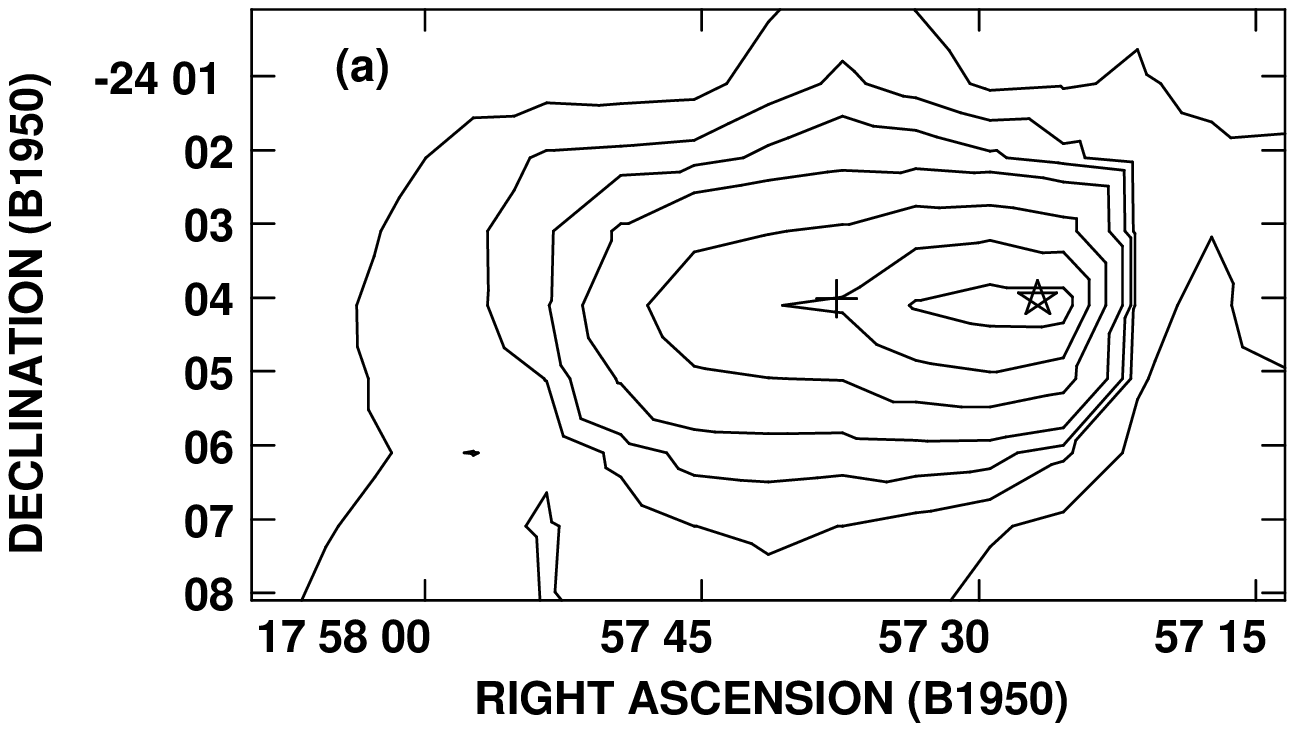}
\vskip -0cm
\end{figure}
 

\begin{figure}
\vskip -6cm
\figurenum{10}
\epsscale{0.7}
\plotone{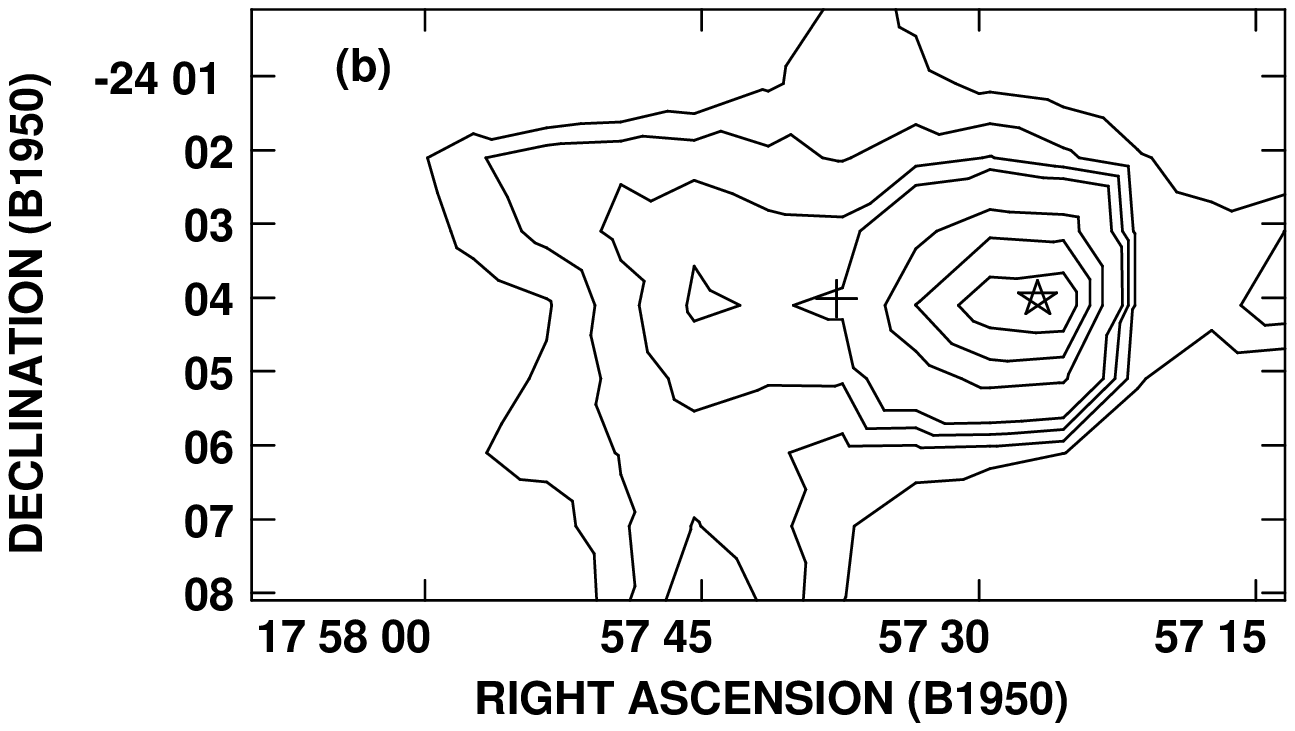}
\vskip -0cm
\figcaption[figure10.ps]{ 
G5.89$-$0.39.
($a$) \hrrl\ line integrated intensity map. Contour levels are 10, 15, 20,
30, 50, 70, and 90\% of the peak value, 5.5~K~\kms. Star and cross
indicate the \uchii\ region and the central compact component, respectively.
($b$) \hrrl\ equivalent line width ($\int T_{\rm L} dv$/\tl) map.
Contours correspond to 40, 45, 50, 55, 70, 80,
and 90\% of the peak value, 47.0~\kms.
}
\end{figure}
 
\clearpage

\begin{figure}
\vskip -1cm
\figurenum{11}
\epsscale{0.7}
\plotone{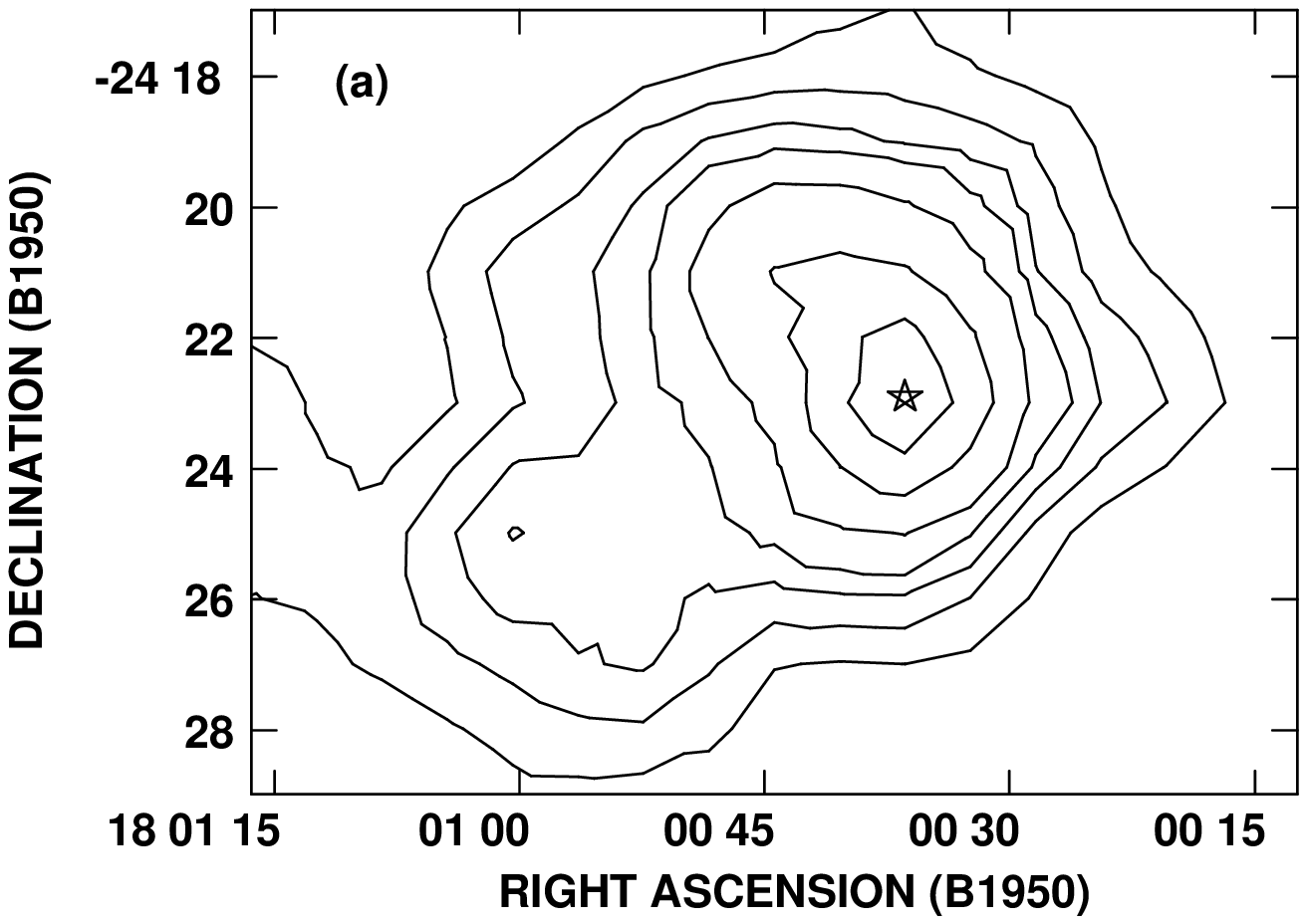}
\vskip -0cm
\end{figure}
 

\begin{figure}
\vskip -3cm
\figurenum{11}
\epsscale{0.7}
\plotone{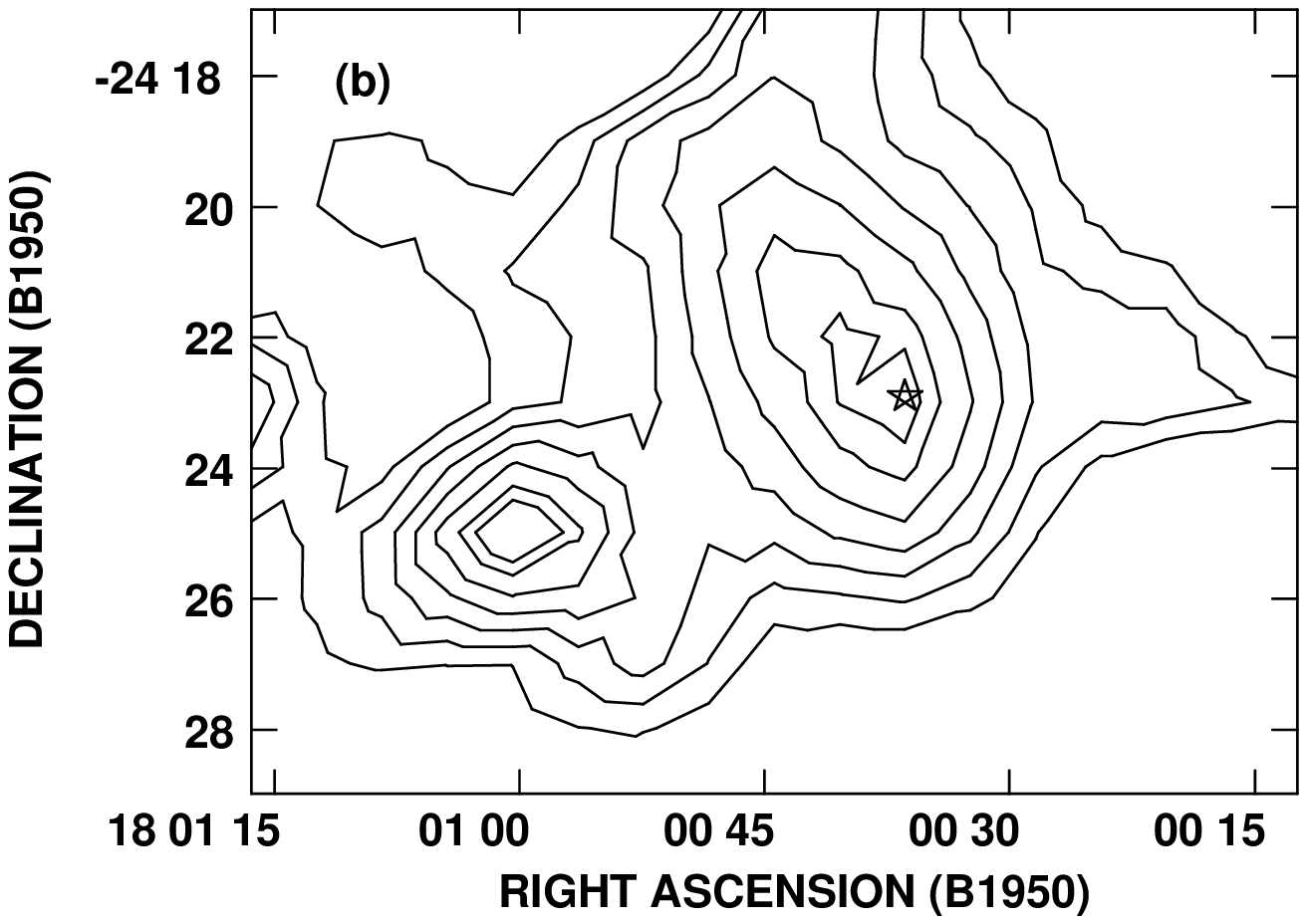}
\vskip -0cm
\figcaption[figure11.ps]{
G5.97$-$1.17.
($a$) \hrrl\ line integrated intensity map. Contours are 15, 20, 25, 30, 40,
60, and 80\% of the peak value, 7.9~K~\kms. Star marks the position of
\uchii\ region.
($b$) \hrrl\ equivalent line width ($\int T_{\rm L} dv$/\tl) map.
Contours correspond to 65, 70, 75, 80, 85, 90, and 93\% of the peak value,
29.6~\kms.
}
\end{figure}


\end{document}